\documentclass[preprint, 12pt ,Fryer 2015, 3p, twocolumn]{elsarticle}
\usepackage{natbib}

\usepackage{epsfig}

\usepackage{amssymb}

\journal{High Energy Density Physics}

\begin{document}

\begin{frontmatter}



\title{Designing Radiation Transport Tests: Simulation-Driven
  Uncertainty-Quantification of the COAX Temperature Diagnostic}


\author{C. L. Fryer\corref{cor1}\fnref{fn1,fn2}}
\author{A. Diaw\fnref{fn1}, C.J. Fontes\fnref{fn1}, A.L. Hungerford\fnref{fn1}, 
  J. Kline\fnref{fn1}, H. Johns\fnref{fn1}, N. E. Lanier\fnref{fn1}, S. Wood\fnref{fn1},
  T. Urbatsch\fnref{fn1}}
\address{Los Alamos National Laboratory, Los Alamos, NM 87545}

\cortext[cor1]{Corresponding Author, fryer@lanl.gov}

\fntext[fn1]{Los Alamos National Laboratory, 
Los Alamos, NM 87545}
\fntext[fn2]{The George Washington University, Washington, DC 20052, USA}

\begin{abstract}

One of the difficulties in developing accurate numerical models of
radiation flow in a coupled radiation-hydrodynamics setting is
accurately modeling the transmission across a boundary layer.  The
COAX experiment is a platform design to test this transmission
including standard radiograph and flux diagnostics as well as a
temperature diagnostic measuring the population of excitation levels
and ionization states of a dopant embedded within the target material.
Using a broad range of simulations, we study the experimental errors
in this temperature diagnostic.  We conclude with proposed physics
experiments that show features that are much stronger than the
experimental errors and provide the means to study transport models.

\end{abstract}


\end{frontmatter}

\section{Introduction}

Laboratory experiments at national laser facilities are designed to
probe a wide range of physics include plasma physics, turbulence
physics, and radiation flow.  For radiation flow, improving the
physics in question requires extreme precision.  We understand the
equations governing this radiation flow; the Boltzman transport
equation is believed to accurately represent what happens in nature
and we can test our numerical methods in reproducing this equation on
simple problems with existing analytic solutions.  To test beyond
these analytic comparisions, we must design experiments that move
past pure-transport solutions.  One example arises from problems
testing how radiation couples to matter (both opacities, equation of 
state  and
hydrodynamics).  The COAX experiment is designed to test one 
aspect of this problem:  the radiation flow across boundary 
layers.

COAX is the successor to Pleaides, an experiment designed as a first
step in studying radiation flow, using a hohlraum drive to study
supersonic diffusive flow down a single foam tube constrained by a
high opacity wall\cite{guymer15,fryer16}.  This experiment had two
sets of diagnostics: soft X-ray spectral emission of the shock front and a
DANTE detection to measure the emergence of the heat front as it
emerged from the foam.  A surprise in this experiment was that the
simulations systematically predicted shorter breakout times for the
radiation front through the tube than those measured in the
experiment.  Although this could be caused by uncertainties in the
physics (e.g.  most equations of state don't accurately include
excitation states of the electrons in the atoms, altering the specific
heat), it was found that uncertainties in the experiment could also
produce systematic errors\cite{guymer15,fryer16}.  To improve upon
this experiment, both a better undestanding of the initial conditions
and the evolution foam properties (e.g. internal energy) are needed to
discriminate between the different interpretations explaining the
disagreement between experiment and simulation.

The COAX platform will ultimately include a DANTE diagnostic at the
end of the target to measure the breakout time.  This DANTE diagnostic
has been used extensively at Omega and its uncertainties are well
documented\cite{2010RScI...81jE505M,2012RScI...83jE117M}.  The current
design of this platform uses the DANTE detector to measure the
hohlraum temperature and a radiograph diagnostic to measure the shock
position.  However, the primary goal of this paper is to understand
the efficacy of a spectral diagnostic to measure the foam temperature
in the target.  This paper will focus on the sensitivities of this
spectral diagnostic.

The spectral diagnostic uses X-ray framing cameries coupled to a
four-strip micro-channel plate detector\cite{2010APS..DPPNP9159T} with
data collected via film with a 0.002\,cm/pixel resolution.  Imaging
absorption spectroscopy makes it possible to resolve gradients in the
material tempurature along the direction of the flow.

The basic design of the COAX experiment includes hohlraum with a laser
entrance on one side that drives a radiation flow on the opposite
side.  The hohlraum is 0.12\,cm in height and 0.16\,cm in diameter
(outer extent) with 0.0025\,cm thick walls.  The setup uses 13 beams
from Omega each delivering 500\,J in a 1\,ns square pulse with
pointings designed to produce a uniform drive through an opening
0.08\,cm in diameter into a target.  The drive through the hohlraum
opening evolves with time, peaking with an effective radiation
temperature of 170\,eV.

\begin{figure}[!hbtp]
\centering
\includegraphics[width=3.0in,angle=0]{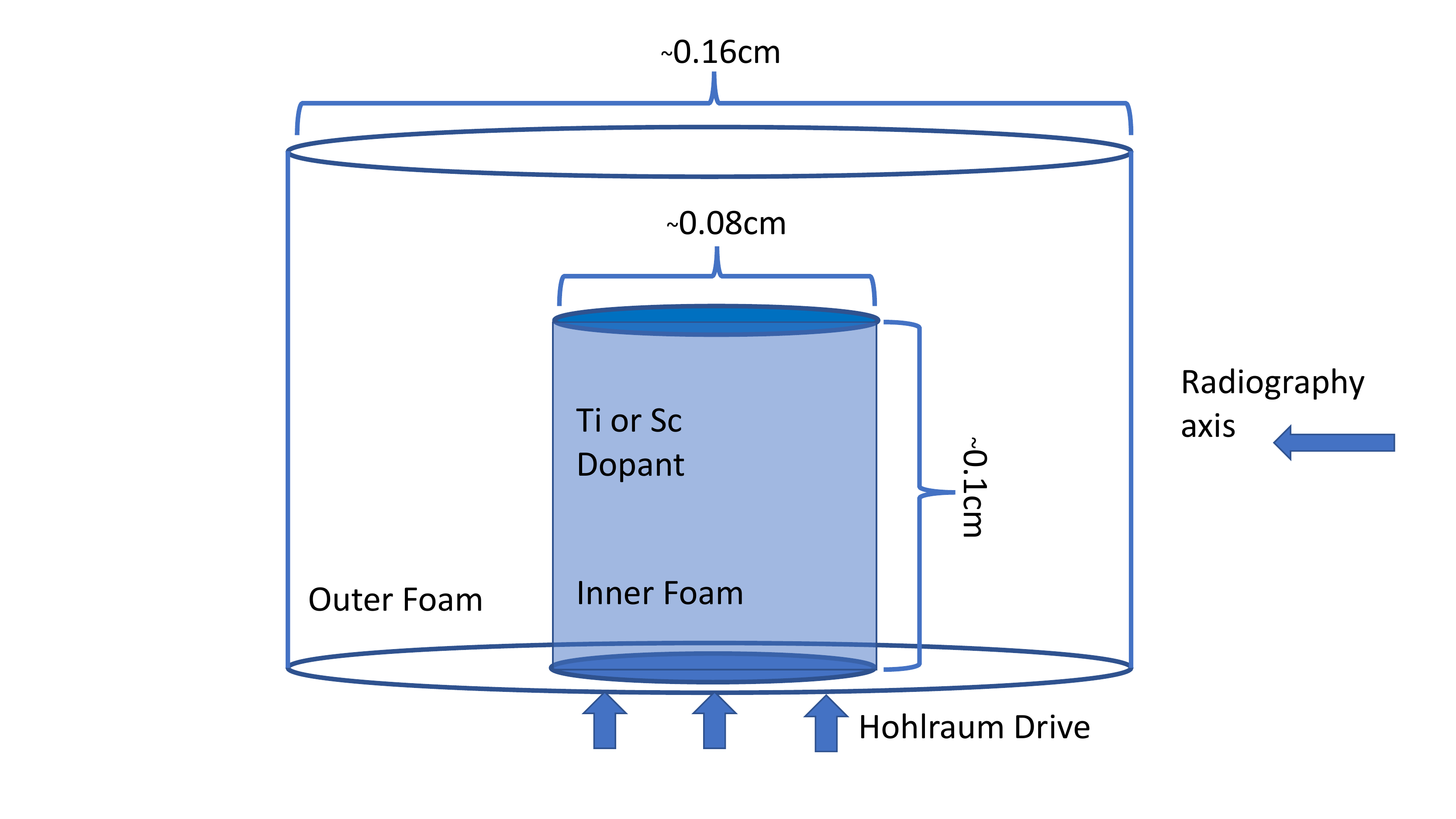}
\caption{Diagram of the COAX experiment with a two-layer target consisting 
of an outer and inner foam above a hohlraum drive.  The new diagnostic 
developed in COAX uses titanium or scandium dopants whose spectral signature 
can be used to measure the temperature of the target.  These can be placed in either the 
inner or outer foams to probe the temperature of the foam and, hence, 
map out the strength of the radiation front across the foam boundary.}
\label{fig:coaxdes}
\end{figure}

An Aluminum foil (with a thickness of 0.0025\,cm) separates the
hohlraum from the target to filter out high-energy photons, providing
a drive that is closer to the blackbody.  The target is shown in
Figure~\ref{fig:coaxdes} with two nested foams above a hohlraum drive.
The foams are composed of a silicon aerogel with densities near
70\,mg\,cm$^{-3}$.  This foam can be doped with titanium or scandium 
to provide the spectral absorption features to measure the temperature 
(e.g. dopants in the form of $TiO_2$).  We have some control of the 
amount and size of the dopants injected into the foam.  The dopants
can be placed in either the inner or outer layers.  One of the major
sources of error in modern radiation transport codes is the treatment
of radiation as it crosses a boundary.  Varying the foam densities
allows experimentalists to alter the boundary conditions to better
probe the radiation physics.  The experiment and its initial results 
will be discussed in detail in a later paper (Johns et al., in preparation).

In this paper, we use a suite of simulations to study the sensitivity of
this diagnostic to the experimental uncertainties.  In particular, we
study the uncertainties of both the initial conditions and the
diagnostic itself in the temperature measurement of the diagnostic.
Section~\ref{sec:init} discusses uncertainties in the initial
conditions including uncertainties in both the drive from the hohlraum
and the densities in the foam.   Section~\ref{sec:diag} discusses the
potential issues with the COAX diagnostic.  As we shall see, the
diagnostic is fairly insensitive to many uncertainties, but there are
some issues that must be understood to fully take advantage of this
diagnostic.  We conclude with a discussion of the future of the COAX
experiment and its ability, given the diagnostic uncertainties, to
address key problems in radiation-flow physics.

\section{Uncertainties in the Initial Conditions}
\label{sec:init}

\begin{figure}[!hbtp]
\centering
\includegraphics[width=3.0in,angle=0]{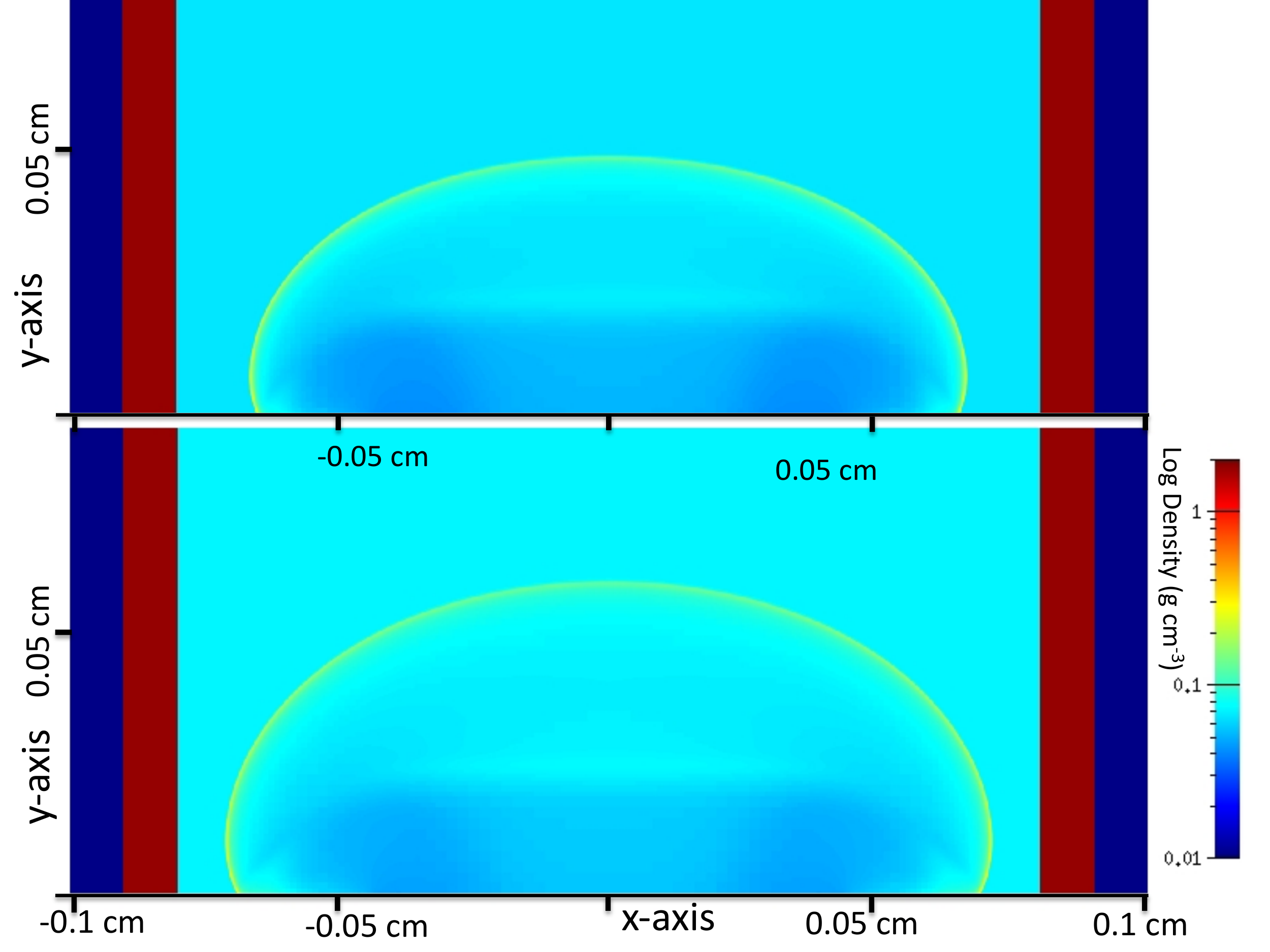}
\caption{Density map of two simulations of the COAX experiment
  2.5\,ns after the launch of the drive, varying the density between
  the upper and lower limits of our 1-$\sigma$ error bars.  }
\label{fig:sample}
\end{figure}

As with many (if not all) laboratory experiments, uncertainties in the
initial conditions can affect the interpretation of the experiment,
limiting the ability of these experiments to constrain the physics
studied in the experiment.  The initial-condition uncertainties were
studied in detail in the Pleiades experiment, focusing on the effect
of these uncertainties on the position of the shock as a function of
time\cite{fryer16}.  The COAX temperature diagnostic provides an
additional constraint to better understand radiation flow
uncertainties.  But, as with the Pleiades experiment, we must first
understand the level of the uncertainties in the initial conditions on
the interpretation of the results from this diagnostic.  Like
Pleiades, the COAX experiment follows the flow of radiation emitted
from a hohlraum through a foam target (Figure ~\ref{fig:sample}).  The
initial-condition uncertainties can be separated into uncertainties in
the characteristics of the target, e.g. the foam characteristics
(density, composition and inhomogeneities) and uncertainties in the
drive (both electron and photons) into this target from the hohlraum
(spectrum, angular distribution, luminosity).

Some aspects of this uncertaities studied in the Pleiades experiment
have improved in COAX, reducing these associated errors.  In addition,
instead of studying the uncertainties in the breakout shock, this
paper focuses on the uncertainties in the temperature diagnostic
developed in COAX.  Understanding the effect of these uncertainties is
critical in determining what physics we can study with COAX.  Although
we ultimately want to use the experiments to test the physics
implementations in the code, in this paper we use these same codes to
determine the extent of the errors from the initial condition
uncertainties.  For this paper, we use the LANL-ASC code Cassio which
includes both implicit Monte Carlo (IMC) and $S_N$ discrete ordinate
methods for the radiation transport coupled onto an Eulerian adaptive
mesh refinement scheme\cite{gittings08}.

This study leverages a grid of simulations probing the different
uncertainties discussed in this paper.  These grids include a set of
models that varied the density and the drive using 9 different foam
densites ranging from $56-77\,{\rm mg \, cm^{-3}}$, 3 drive powers
varying the power by $\pm$5\%, 2 durations varying the duration of the
drive by $\pm$10\%, 2 angles and 2 spectral distributions.  We run
these models with both Implicit Monte Carlo and $S_N$ options, leading
to a total of over a four hundred models in our base grid.  We ran a
suite of simulations of both angle and energy in the drive for a
focused study on these effects (another 20 models).  We also ran 20
models varying the foam inhomogeneities (both the density variations
and the positions of the inhomogeneities).  We ran another suite
studying the dopant size (another 20 calculations).  In total, this
work summarizes the results of over 500 caclulations.

\subsection{Drive}
\label{sec:drive}

The laser drive itself is generally well-understood with errors less
than $\pm 2-3$\%.  However, modeling the hohlraum is difficult for
codes using the Euler equations as the basis for the hydrodynamics.
The Euler equations for momentum, mass and energy conservation
implicitly assume matter is in thermal equilibrium.  In the
low-density conditions of the hohlraum, not only are the ions,
electrons and photons decoupled with respect to each other, but these
particles tend not to be in equilibrium in parts of the
hohlraum\citep{renderknecht18,Brodrick17,Sherlock17}.  

In the hohlraum, such time-independent assumptions are not satisfied,
and the calculations are suspect.  Simulations can provide a
reasonable estimate of the equilibration timescales for different
quantities (see appendix).  Given the large mean free path of the
electrons in both the hohlraum and the target, it is not clear that
the electron distribution is a Maxwellian and can be described by a
single temperature.  If we nonetheless assume the distributions of the ions and
the electrons can be described by a Maxwellian, we can then determine
the coupling between these two species.  We find that the
equilibration timescales at $300\,eV$ for electron-ion coupling within
the hohlraum and a density of air lie between 1-2\,ns and, for
near-vacuum densities of $0.03 \,{\rm mg \, cm^{-3}}$, $100$\,ns.  The
corresponding timescales for the major atomic level states to be
within 10\% of their time-independent value are on the order of a few
ns.  Given that the timescales for our COAX experiment is only a few
ns, our series of equilibrium and time-independent assumptions is
highly suspect for our hohlraum modeling.  To date, the primary HEDP
codes include minor corrections for out-of-equilibrium physics and
time-dependent effects are not included in the atomic level states.
This means that most hohlraum models are approximations at best.  It
is because of these limitations that we have a rather large
uncertainty in our drive.  The Euler equations used in codes like
Hydra~\citep{marinak01}, RAGE\citep{gittings08} or
FLAG\citep{harrison10} do not capture this out-of-equlibrium physics
to model the hohlraum and, although they can be used to make a first
order estimate of the emission, simulations using these codes can lead
to incorrect estimates of averaged terms like
heating\citep{Brodrick17}.

\begin{figure}[!hbtp]
\centering
\includegraphics[width=3.0in,angle=0]{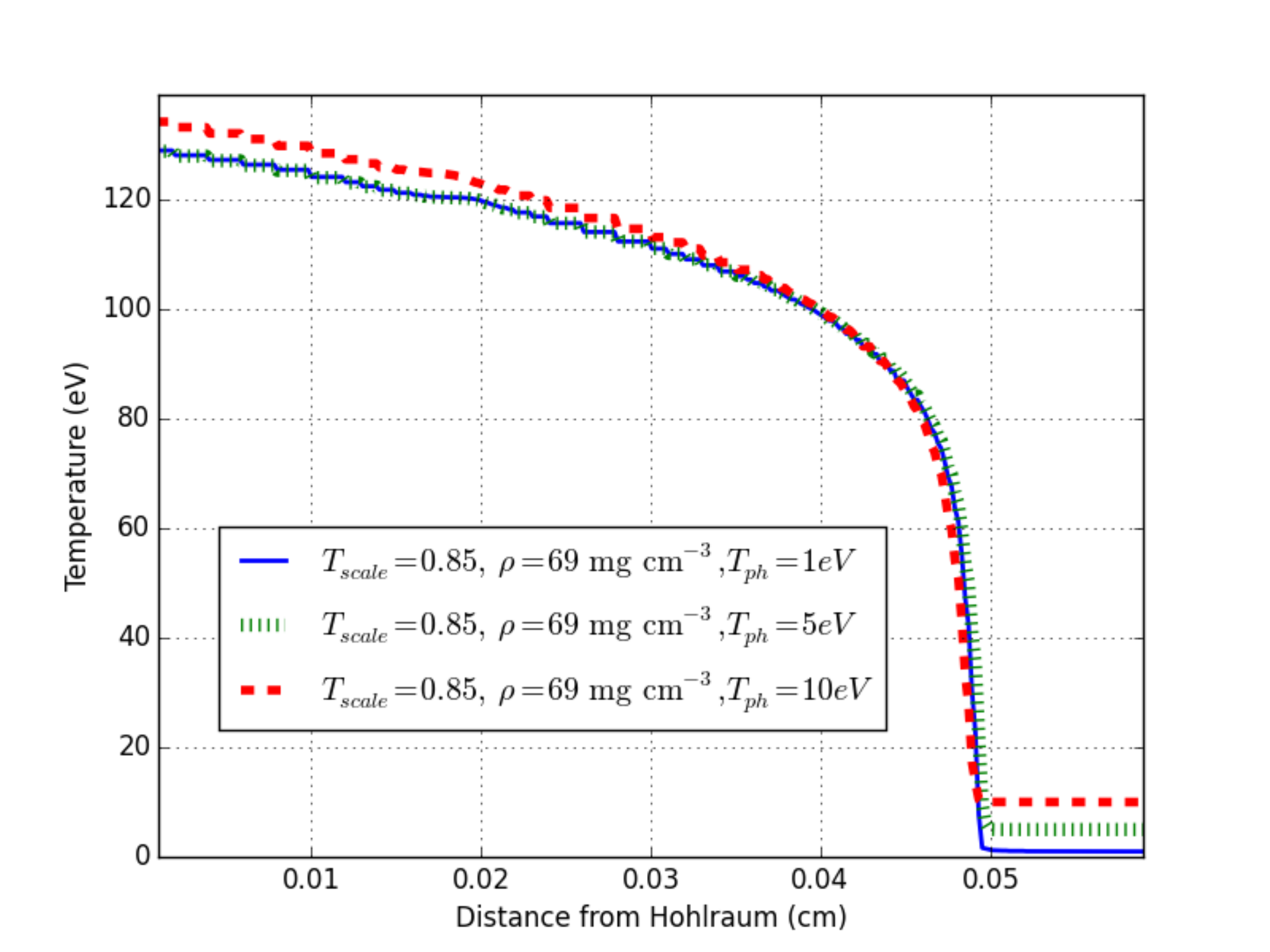}
\caption{Temperature profile versus distance from the hohlraum drive
  (0.01\,cm from the axis) of our standard drive and density target at
  2\,ns.  We have varied the electron preheat from 1 to 10\,eV.  This
  preheat does not change the position of the shock and only a 10\,eV
  preheat makes a noticeable ($1-2$\,eV) change in the temperature
  profile.}
\label{fig:preheat}
\end{figure}

Without detailed measurements of the drive coming out of the hohlraum,
we are limited to using these first order estimates.  To better
understand the drive uncertainties, we have conducted a number of
studies of electron heating, photon flux and spectrum.  Non-thermal
electrons produced in the hohlraum can stream into the target,
preheating the target material prior to the launch of the radiation
front.  For low-drive experiments, a few eV pre-heat can alter the
evolution of the shock front\citep{falk18}.  But, for the COAX
experiment, even a 10\,eV pre-heat does not significantly alter the
flow of the much more powerful COAX drive.  Figure~\ref{fig:preheat}
shows the temperature profile of 3 separate simulations at 2\,ns with
a 1, 5, and 10\,eV preheat.  The 10\,eV preheat (larger than expected
from past studies\citep{falk18}), produces only a small ($\sim
1-2$\,eV) difference in the profile temperature.  Our analysis
supports the practice in most radiation transport experiments that
this heating is unimportant because both the amount of high-energy
emission (electrons and photons) is small compared to the total
emission and most of this emission streams through the target.

We have also varied the spectrum and angular distribution of the
radiation flowing out of the hohlraum.  If all of the photons are
non-thermal with temperatures above 1\,keV, they can preheat the
target, altering the flow.  But for a wide range of fluxes (pushing
the spectrum above 200\,eV or down below 100\,eV), the effect is less
than a few percent in the shock position and less than 1-2\,eV in the
temperature profile.  Similarly, we produced only small differences
when we altered the angular distribution of the radiation flux from a
typical Lambertion to a forward-peaked function: $f(\theta) \propto
cos^5(\theta)$.  For our drive, the target is sufficiently optically
thick to reset the spectral and angular features.

\begin{figure}[!hbtp]
\centering
\includegraphics[width=3.0in,angle=0]{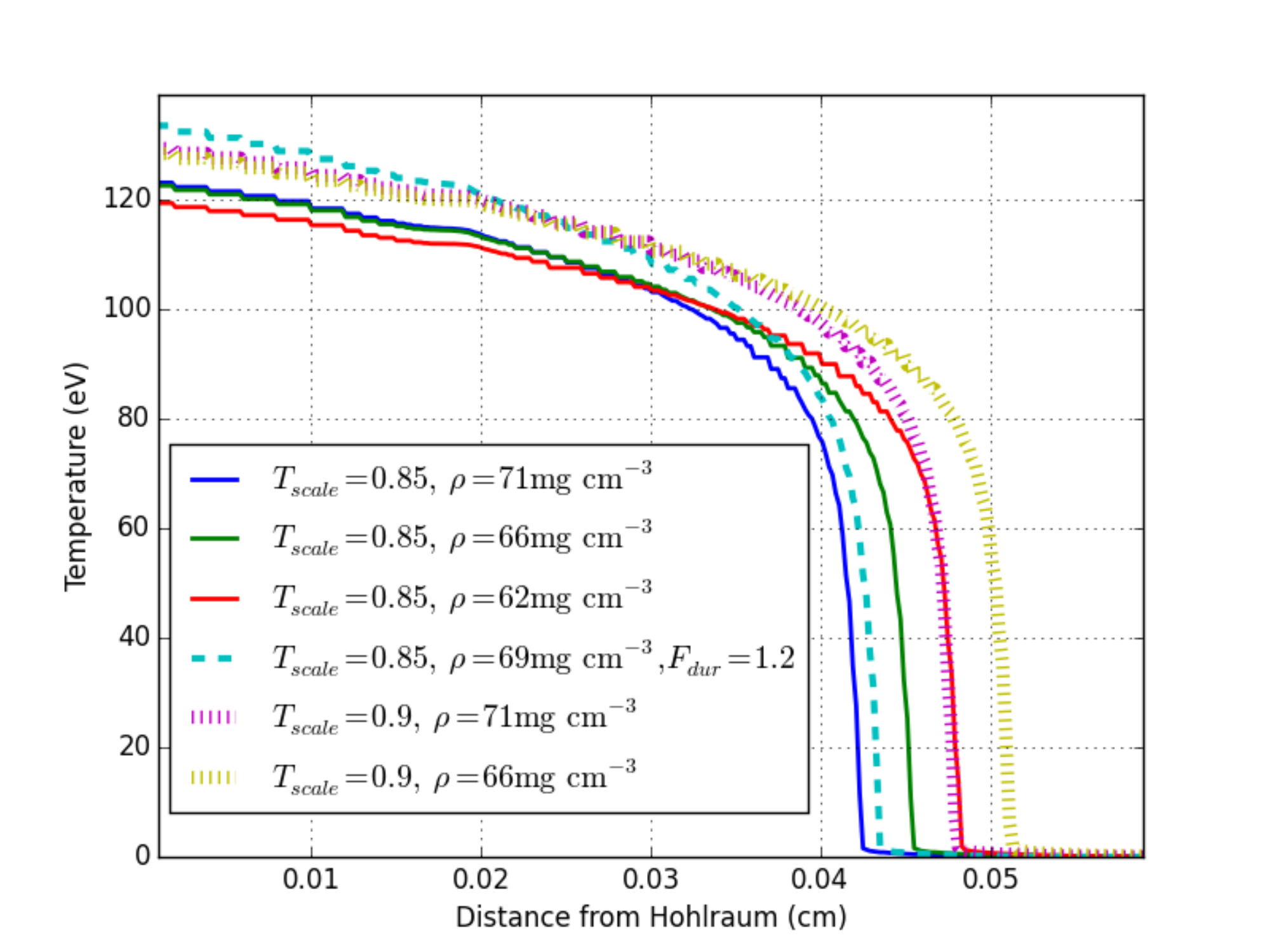}
\caption{Temperature profile versus distance from the hohlraum drive
  (0.01\,cm from the axis) at 2\,ns assuming no preheat and varying
  both the drive characteristics (duration, power) and foam density.
  The uncertainties in the drive characteristics lead to an
  uncertainty in the temperature profile of $\pm 8$\,eV.  These
  variations also alter the position of the shock.  However, the
  measured uncertainties in the foam densities can drastically alter
  the shock position.}
\label{fig:comptemp}
\end{figure}

Finally, we varied the power and duration of the radiation emitted
from the hohlraum.  Figure~\ref{fig:comptemp} shows the results for a
subset of our grid of calculations where we scale the temperature of
the drive by 5\% (varying the flux by plus or minus 10\%), the
duration of the drive by 20\% (altering the total energy by plus or
minus 10\%) and the foam density.  The change in power can alter the
position of the shock at 2\,ns by 0.007\,cm or over 10\%.  The drive
duration doesn't alter the forward shock position much but it can
alter the temperature behind the shock which can dominate what we see
in the diagnostic measurement.  The power and duration alter the
temperature by $\pm 5$\,eV and $\pm 7$\,eV respectively.  Combined,
the effect is over $\pm 8$\,eV.  With no constraint on the shock
position, the uncertainties in our drive lead to uncertainties in the
temperature behind the shock of $\pm 8$\,eV.  Ideally, we can use the
shock position to further constrain the drive.  We will discuss this
further in our description of the target uncertainties
(Sec.~\ref{sec:target}).

\subsection{Target Uncertainties}
\label{sec:target}

In the Pleiades experiment, foam density measurements typically had
3-4\% errors and different measurement techniques produced different
results\cite{fryer16}.  To study the dependence on density, we
varied the density for our COAX experiment from $62-71\,{\rm mg \,
  cm^{-3}}$.  By combining these variations with our drive, we can
determine the full errors from our initial conditions.  Typically, the
density variations alter the position of the shock at a given time
with a slightly more modest effect on the temperature profile.  In
Section~\ref{sec:drive}, we pointed out that we might be able to use
the shock position to constrain the uncertainties in the drive.
However, density uncertainties alter the shock position in a
comparable way to the drive uncertainties.  With a shock position, we
can constrain the drive/density pair, but the uncertainty in the
absolute temperature profile remains $\pm 8$\,eV.  We note, however, 
that density uncertainties are believed to be smaller for silicon 
foams and this may overestimate the errors caused by density uncertainties.

\begin{figure}[!hbtp]
\centering
\includegraphics[width=3.0in,angle=0]{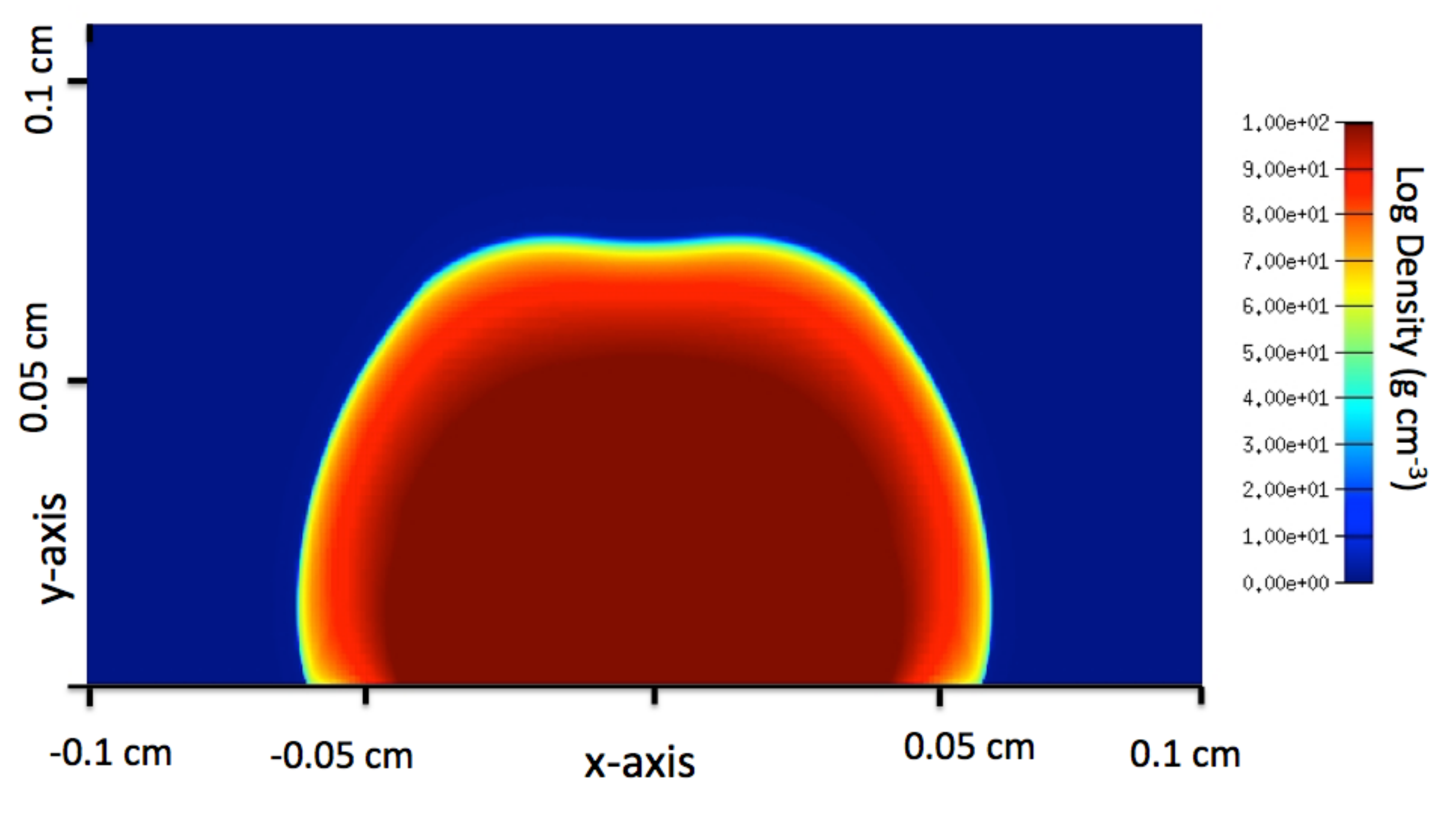}
\caption{Density map of a COAX experiment simulation 2.0\,ns 
after launch of the drive including inhomogeneities into the 
target foam but fixing the average foam density.  The inhomogeneities 
can alter the shock front some, but the effect depends sensitively on 
the placement of the inhomogeneities.}
\label{fig:inhomimage}
\end{figure}

\begin{figure}[!hbtp]
\centering
\includegraphics[width=3.0in,angle=0]{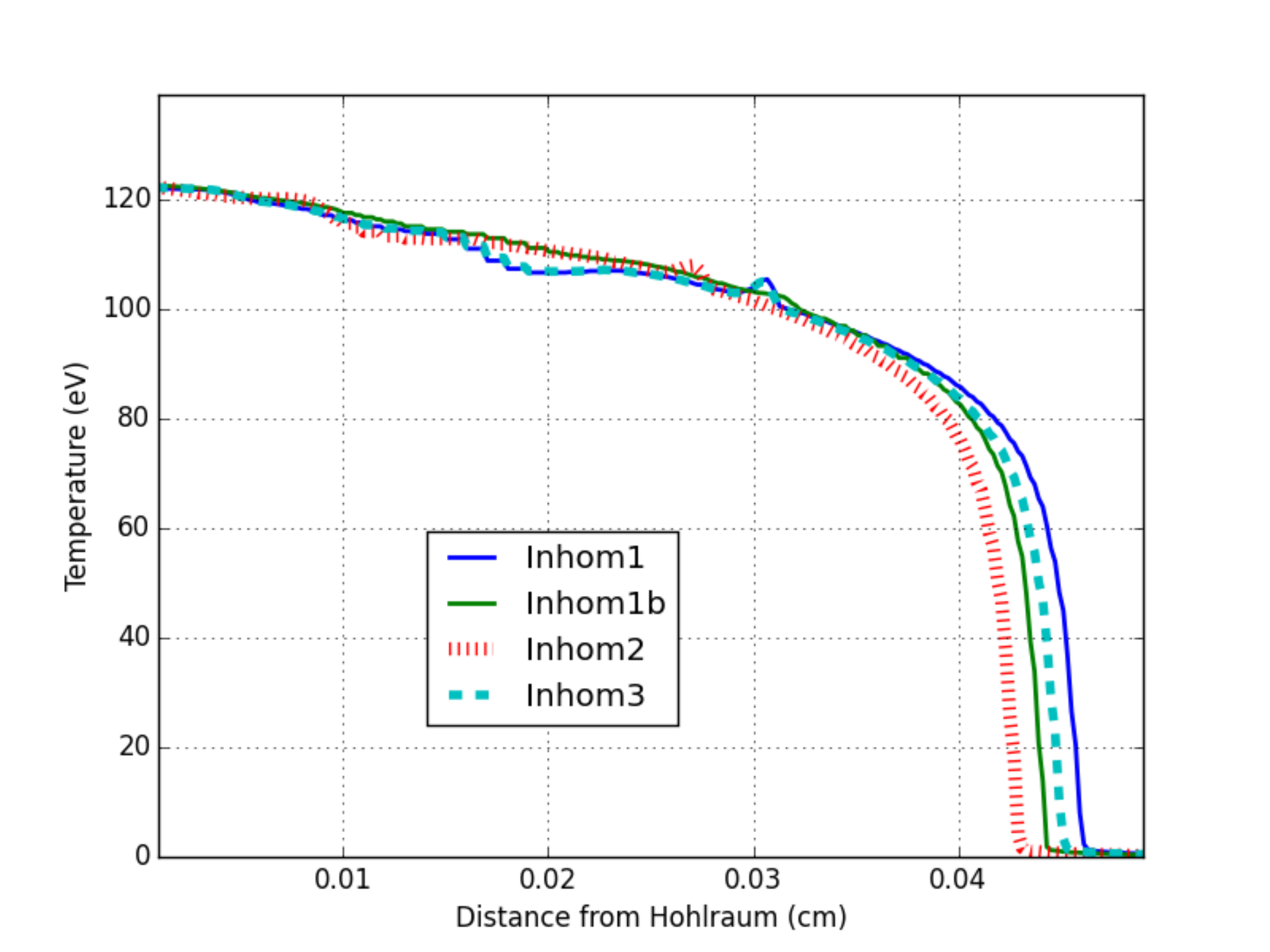}
\caption{Temperature profile versus distance from the hohlraum drive
  (0.01\,cm from the axis) at 2\,ns assuming no preheat for 3
  different instantiations of the density inhomogeneities where we
  varied the geometry of the perturbations but kept the average
  density constant (Inhom1, Inhom2, Inhom3) with 50\% enhancements at
  the centroids of the perturbation.  Inhom1b shows a model where the
  perturbations are 10\% enhancements.  Inhomogeneities can alter the
  shock position by 10-20\% and can speed up or slow down the shock.}
\label{fig:inhomtemp}
\end{figure}

In addition to uncertainties in the average foam density, detailed
studies of the Pleiades experiment also found that most foams had
fairly large scale inhomogeneities in them\cite{fryer16}.  For the
COAX experiment, we introduce a series of clumps into the foam,
keeping the average density constant.  Figure~\ref{fig:inhomimage}
shows the density map of a simulation including inhomogeneities.  The
placement of the inhomogeneites affects both how much the
inhomogeneities alter the density map and the arrival times.
Figure~\ref{fig:inhomtemp} shows our standard temperature profile map
using four different instantiations (different distributions of the
clumping) of the density inhomogeneities.  Inhomogeneities do not
affect the temperature dramatically, but can alter the position of the
shock by 10-20\%.

With the imaging spectroscopy in COAX, we can not only measure the
temperature in the radiation flow, but its slope.  For the 6 models
presented in figure~\ref{fig:comptemp}, the slopes are: -674, -615,
-530, -816, -596, -597 $eV \, cm^{-1}$ respectively for the 4 $T_{\rm
  scale}=0.85$ models ($\rho = 77, 66, 62 {\rm \, g cm^{-3}}$ plus the
$\rho = 69 {\rm \, g cm^{-3}}, F_{\rm dur}=1.2$ model) and the two
$T_{\rm scale}=0.9$ models ($\rho = 69 , 62 {\rm \, g cm^{-3}}$).
This slope can be used to further distinguish between models.  For
instance, although the temperatures are very similar between the
$T_{\rm scale}=0.85$, $\rho = 69 {\rm \, g cm^{-3}}$, $F_{dur}=1.2$ model
and the $T_{\rm scale}=0.9$, $\rho = 71 {\rm \, g cm^{-3}}$ model, the 
temperature gradients are very different.  The slope doesn't distinguish 
between all uncertainties (e.g. the effects of inhomogeneities), but it 
does provide additional constraints.

Based on the modeling of the Pleiades experiment, these uncertainties
are what we expect from foam-based targets and unmeasured hohlraum
drives.  By including measurements of the shock position, we can
constrain some of the degeneracies in our errors, but we can not
remove the temperature variations caused by the combined
uncertainties.  Given the current uncertainties in the drive and target, we
can not predict the absolute temperature behind the shock to better than $\pm
8$\,eV and, hence, can't confirm this diagnostic's effectiveness to
better than $\pm 8$\,eV.  Decreasing the errors in the drive and 
foam target properties can reduce this uncertainty.  In addition, the 
errors in the relative changes with this diagnostic may be much smaller.  
To understand this, we must understand the diagnostic uncertainties.

\section{Temperature Diagnostic Uncertainties}
\label{sec:diag}

The temperature diagnostic of the COAX experiment measures the
absorption line spectra of a dopant backlit by an X-ray source.  As we
learned in Section~\ref{sec:init}, the initial condition uncertainties
in this experiment prevent us from validating this probe of the
temperature to better than $\pm 8$\,eV.  We can, however, use
simulations to better understand the physics uncertainties of this
diagnostic and determine the conditions where it is most effective. 

The temperature of the diagnostic is more complex than a simple 
single-temperature/density profile: the breakdown of equilibrium
assumptions, the uniformity of the dopant, and the uniformity of the
foam target along the line-of-site of our X-ray source.  In many
cases, the deviations caused by these complexities do not affect the
accuracy of this diagnostic.  We review each of these in turn.

\subsection{Equilibration}

Although the Cassio code does not assume strict local thermodynamic
equilbrium, many of its physics components include equilibrium
assumptions.  For example, although the code follows both the ion and
electron distributions separately, both are assumed to follow a
Maxwell-Boltzmann distribution described by a single temperature.  The
radiation field is followed in multi-group (typically 60-100 groups)
following the full Boltzmann transport equation.  The opacities
typically are set by the electron temperature (assuming radiation and
matter equilibrium) in a steady state solution and the time implicit 
transport scheme relies on these.  Different properties 
of the physics are modeled at different levels of fidelity and equilibrium 
states.  It is worth studying, for our COAX problem, where such a code 
is valid.

\begin{figure}[!hbtp]
\centering
\includegraphics[width=3.0in,angle=0]{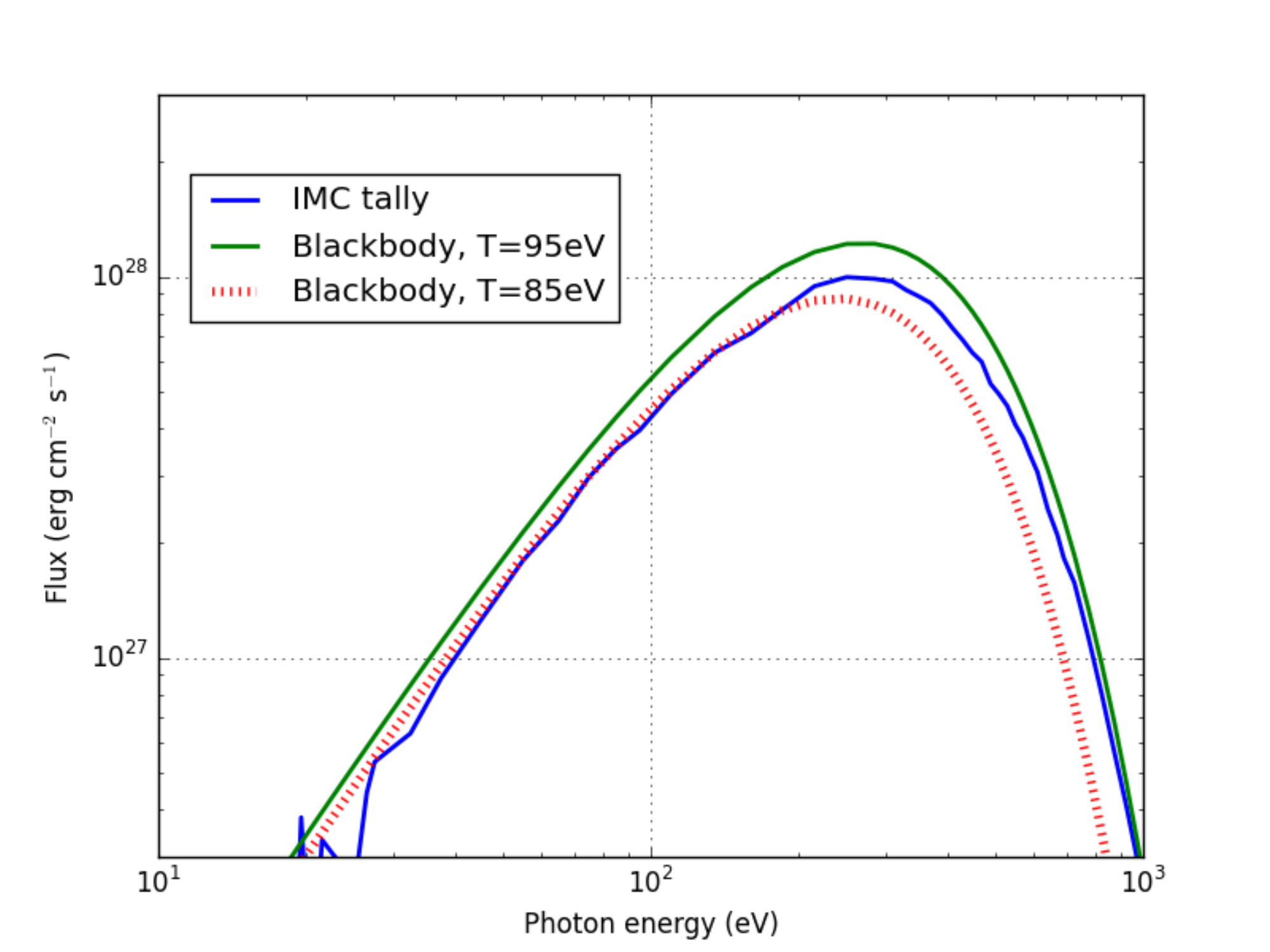}
\caption{Spectrum (flux versus energy) of the radiation front behind
  the shock in our COAX simulation compared to two blackbody solutions.  
  Although the spectrum is not a strict blackbody, the variation is minimal.  
The effective temperature radiation of the radiation is slightly higher 
than the 80-85\,eV matter temperature due to the slight delay in the 
radiative heating of the matter.}
\label{fig:bbcomp}
\end{figure}

As we have already discussed in Section~\ref{sec:init},
time-independent, equilibrium assumptions are not justified for all
aspects of a laser-driven laboratory experiment.  We showed that the
conditions in the hohlraum are insufficient to evolve into equilibrium
on the timescales of the experiment.  But for the higher densities in
the foam target, these assumptions are more valid.  The ion-electron
equilibration timescales are 0.3\,ns for a $60\,{\rm mg \, cm^{-3}},
100\,eV$ foam.  The electron equilibration times are even shorter.
The mean free path of the photons is also low enough for all but the
highest energy photons from the drive that the radiation quickly
equilibrates as well\footnote{Note that we are assuming the electrons
  are described by a Maxwellian and this may speed the equilibration
  of the photons.}.  Figure~\ref{fig:bbcomp} shows the spectrum just
behind the radiation front from a 100 group IMC calculation compared
to two blackbody profiles.  Although the radiation is not a strict
blackbody, it is very close to it.  The corresponding matter
temperature at this time for this tally surface is roughly 80-85\,eV,
slightly below the radiation temperature as it quickly equilibrates to
the radiation.

\begin{figure}[!hbtp]
\centering
\includegraphics[width=2.8in]{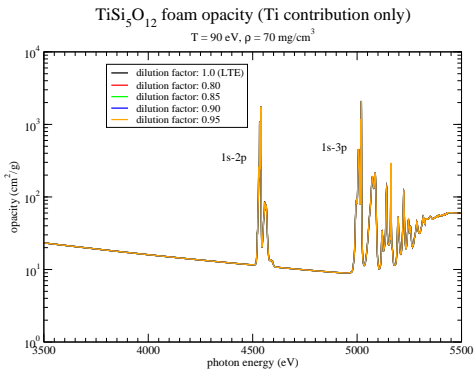}
\caption{Opacity of the titanium dopant comparing a local
  thermodynamic equilibrium (LTE) solution to solutions where the
  radiation flux is reduced or diluted below the LTE temperature.
  This dilution reduces the role radiation plays in setting the level
  states.  Because collisions dominate the level states, it is not
  surprising that the spectrum is not affected by this dilution.  This
  calculation assumes that the radiation is described by a single
  temperature and that the level states are in equilibrium.  Although
  the equilibration time is fast, it is likely that time-dependent
  effects are more important, especially for measurements made right
  after the shock has passed.}
\label{fig:tiequi}
\end{figure}

This small variation, both in the blackbody nature of the radiation
and the equilibration between radiation and electrons, suggests that
the thermodynamic equilibrium can also be assumed for the atomic level
states.  Indeed, at these high densities, the collision rate is
typically high, and the atomic level states reach an equilibrium
solution quickly, the corresponding $\beta$ value measuring the
departure from local thermodynamic equilibrium is less than $\sim
0.02-0.03$ \cite{busquet93}.  Because collisions dominate, any small
differences between the radiation spectra and the electron temperature
will not affect the level states of our dopant (see
Figure~\ref{fig:tiequi}).

\subsection{Dopant Uniformity}

\begin{figure}[!hbtp]
\centering
\includegraphics[width=3.0in]{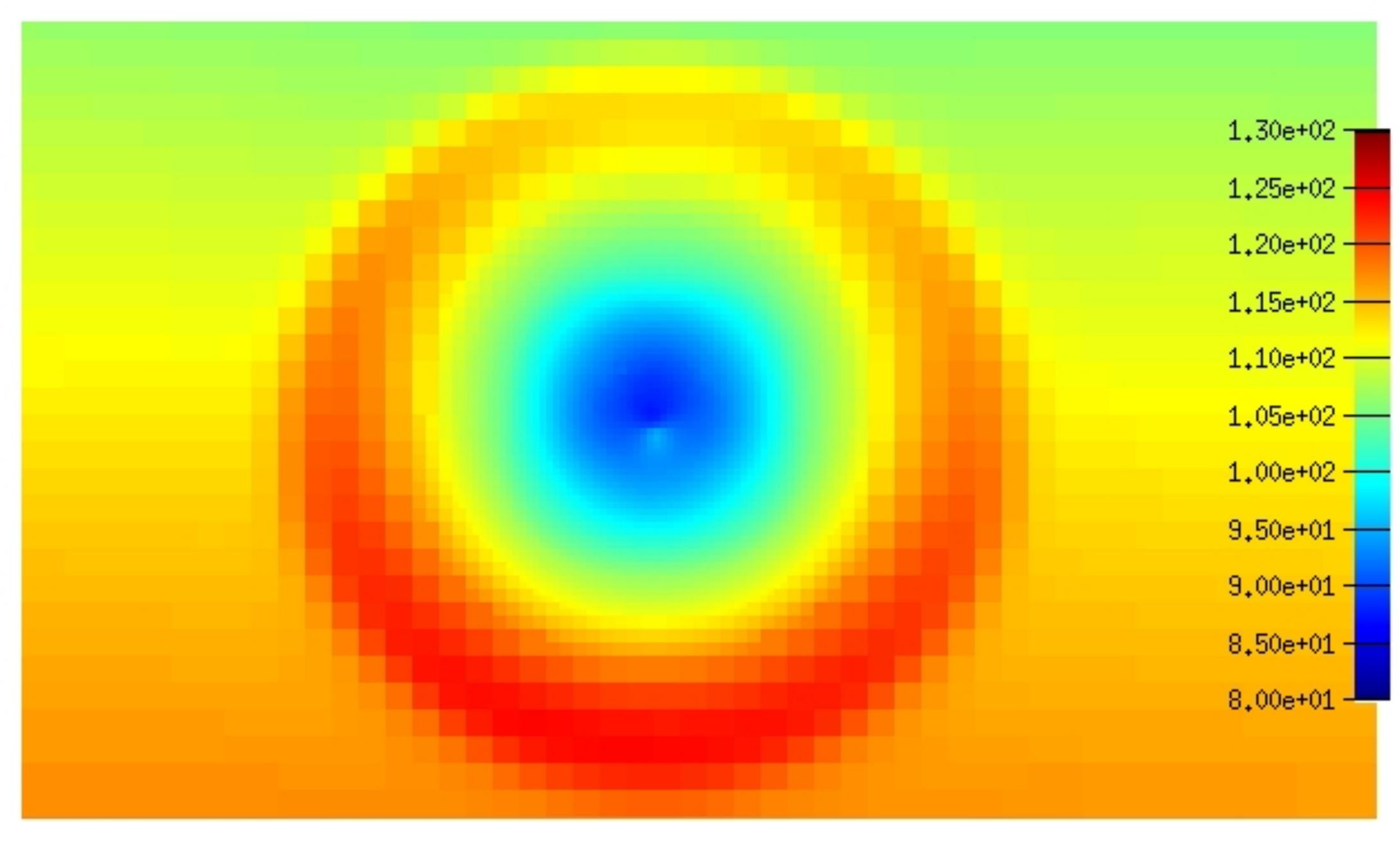}
\includegraphics[width=3.0in]{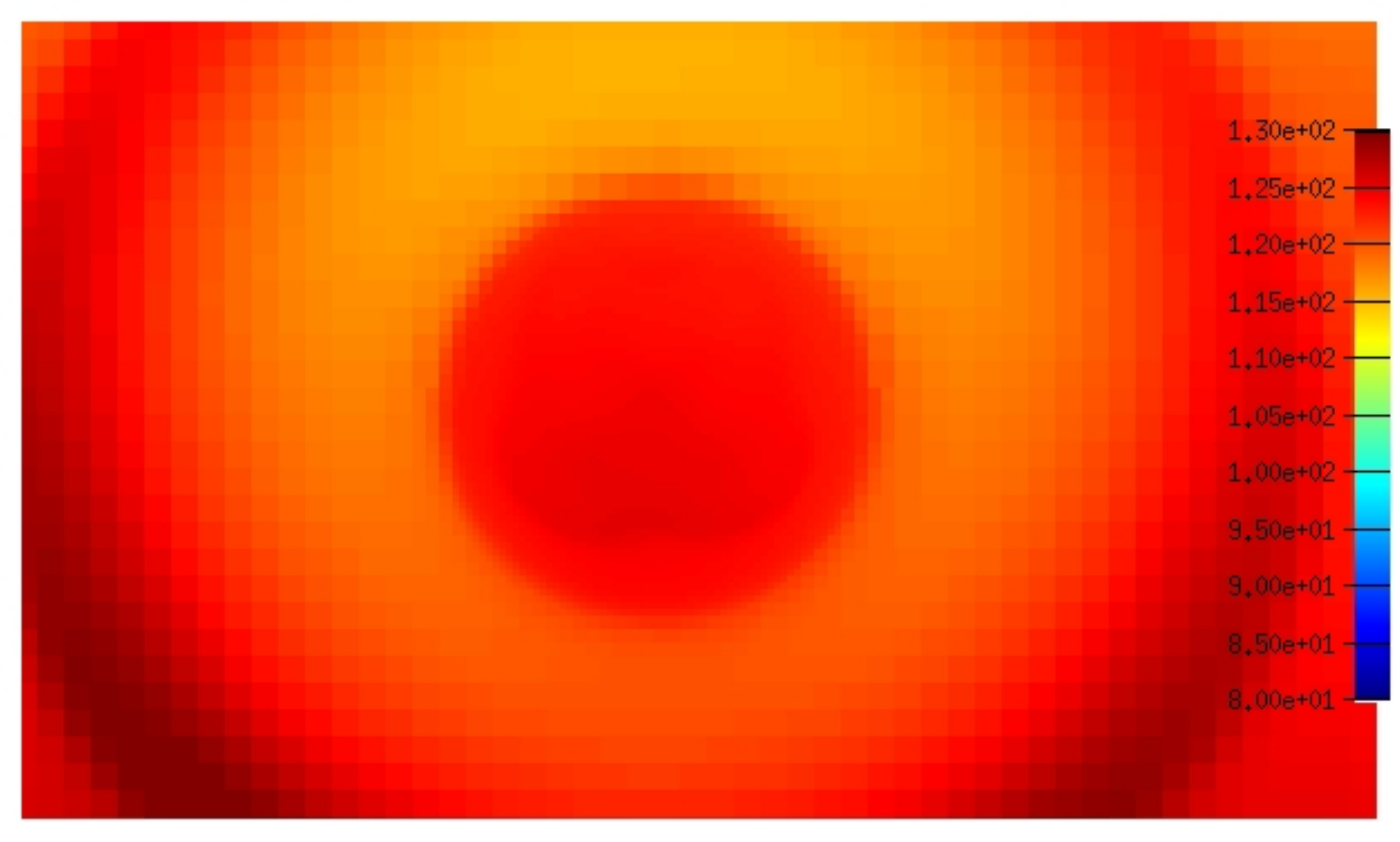}
\caption{Temperature of a $1\,\mu m$ dopant 0.15 and 0.3\,ns after 
the radiative front has crossed the dopant.  By 0.3\,ns, the temperature 
variation across the doped material is less than 10\,eV.}
\label{fig:dop1}
\end{figure}

\begin{figure}[!hbtp]
\centering
\includegraphics[width=3.0in]{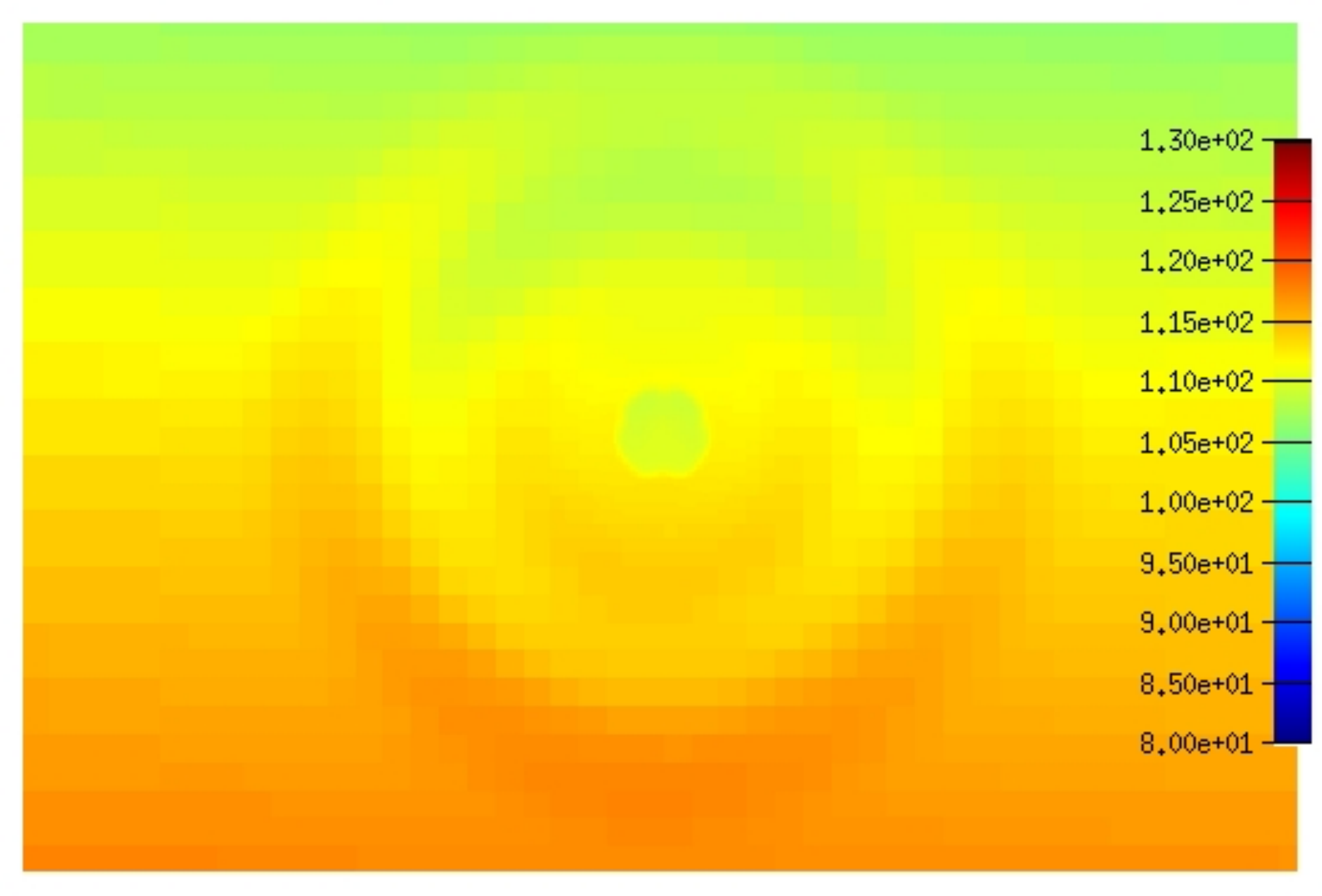}
\includegraphics[width=3.0in]{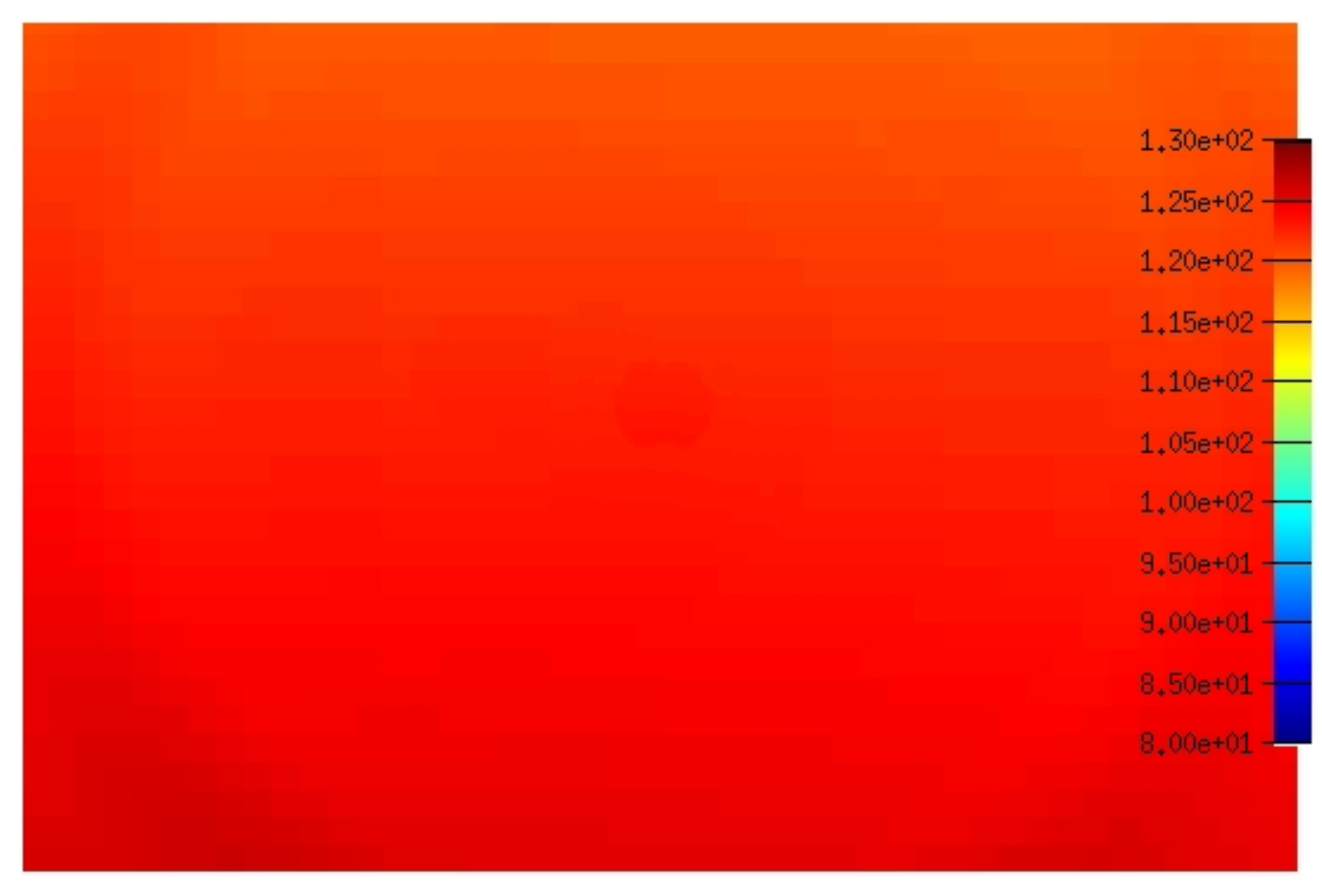}
\caption{Temperature of a $0.3\,\mu m$ dopant 0.15 and 0.3\,ns after 
the radiative front has crossed the dopant.  By 0.3\,ns, the temperature 
variation across the doped material is less than 1\,eV.}
\label{fig:dop2}
\end{figure}

Although the radiation and matter quickly equilibrate in the foam, the
equilibration time of the dopant may take longer.  For sufficiently
large dopant sizes, the timescale to reach an equilibrium temperature
is longer than the experimental timescale.  This means that the dopant
is not necessarily an accurate measure of the temperature of the foam.
To test this for our experiment, we have used the adaptive mesh
refinement capability in Cassio, we are able to resolve the dopants
and determine the timescale for dopants of different size to
equilibrate to a uniform temperature.  For large dopants ($10\, \mu
m$), a portion of the dopant would remain cool ($<20 \, {\rm eV}$)
throughout the duration of the experiment.  Figure~\ref{fig:dop1}
shows the temperature variation for a $1\,\mu m$ dopant 0.15\,ns and
0.3\,ns after the radiation front passes across the dopant.  Although
the variation is large at 0.15\,ns, so is the shock front temperature
gradient.  The effective radiation temperatue and matter temperature
are also not in equilibrium at this time and our opacity calculation
is not valid.  0.3\,ns after the passage of the radiation front, the
variation is only 10\,eV.  Our dopants are typically a fraction of a
micron.  Figure~\ref{fig:dop2} shows the same variation for a
$0.3\,\mu m$ dopant.  In this case, the dopant has equilibrated within
0.3\,ns of the passage of the shock (temperature variations of less
than an eV).  For the COAX experiment as run, the dopant size is not a
source of error.

\subsection{Diagnostic Uniformity}

Even if the dopants heat quickly and integrate with the foam, our diagnostic 
traces through a range of conditions and we do not measure a single density/temperature 
condition.  The radiation front has a curvature associated with it as it expands
laterally (e.g. Figures~\ref{fig:sample},\ref{fig:inhomimage}).  The
nature of this curvature will vary with different transport
prescriptions and these variations must be extracted from the observed 
spectra that integrates across the line of site of our experimental 
diagnostic.  Although the analysis is much more difficult than simply 
assuming a single density/temperature, by understanding and including this variation 
in our analysis, we can construct a much more accurate diagnostic of the radiation 
flow.

\begin{figure}[!hbtp]
\centering
\includegraphics[width=3.0in,angle=0]{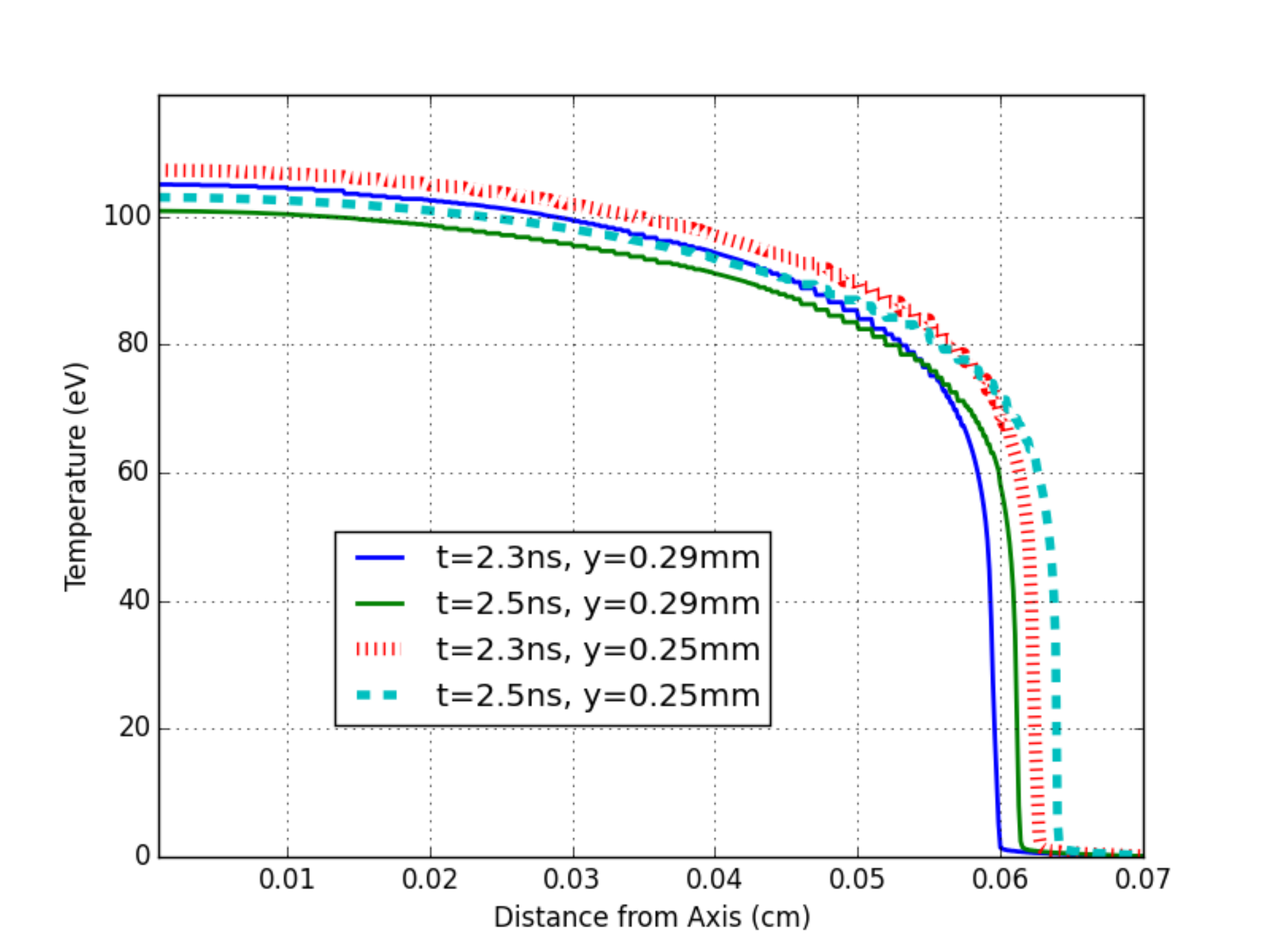}
\caption{Temperature profile along the line-of-site of the diagnostic
  X-rays at two different positions ($0.025,0.029$\,cm above the
  drive) in our target at 2 different times ($2.3,2.5$\,ns).  Within
  the doped region, the the temperature varies by over 4\,eV.  The
  X-ray diagnostic is actually a slit $30 \, \mu m$ wide that takes data
  over a 0.1-0.2\,ns window.  This further broadens the range of
  temperatures observed in a single data point to $\pm 5$\,eV.}

\label{fig:tracetemp}
\end{figure}

\begin{figure}[!hbtp]
\centering
\includegraphics[width=3.0in,angle=0]{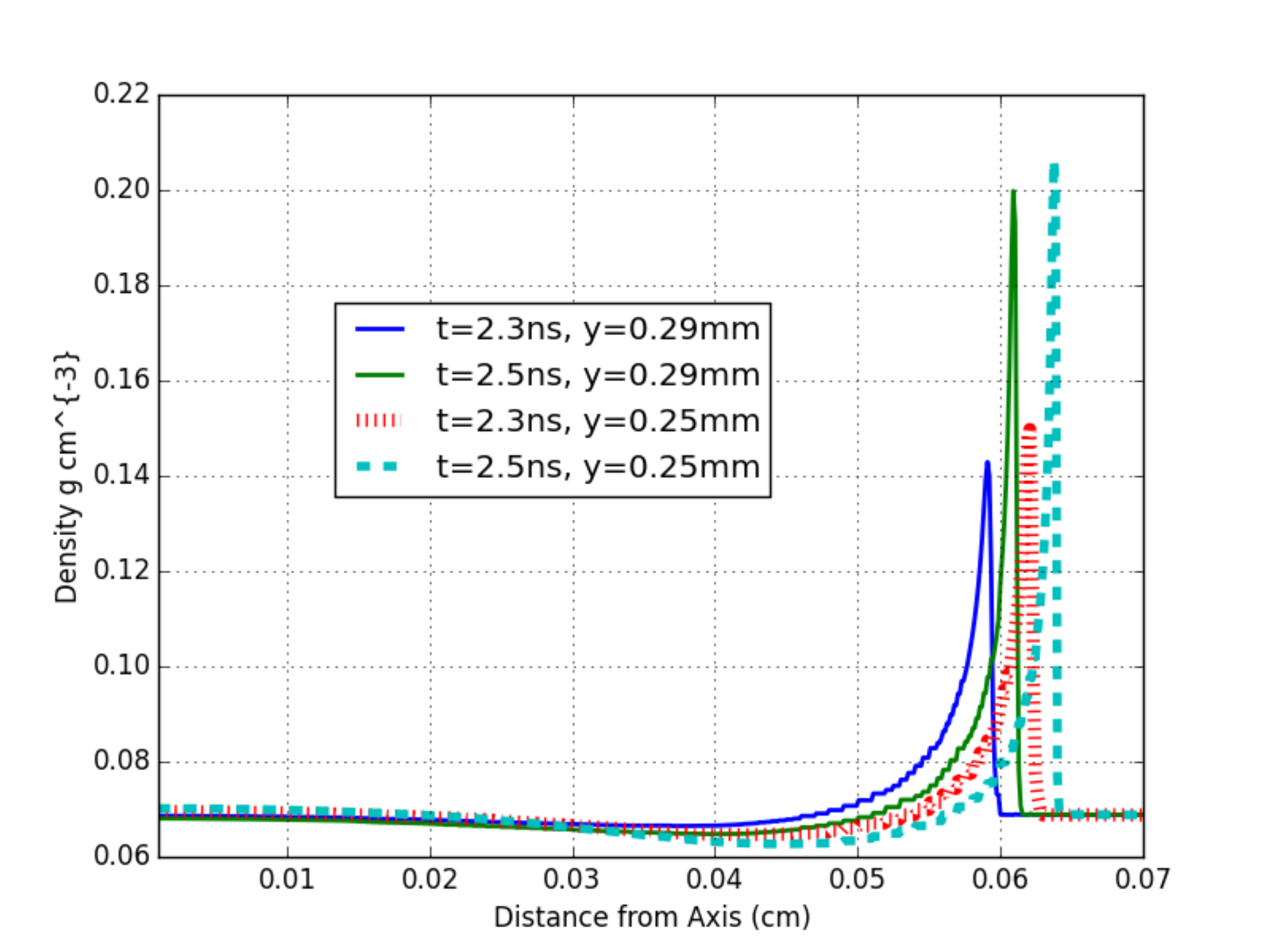}
\includegraphics[width=3.0in,angle=0]{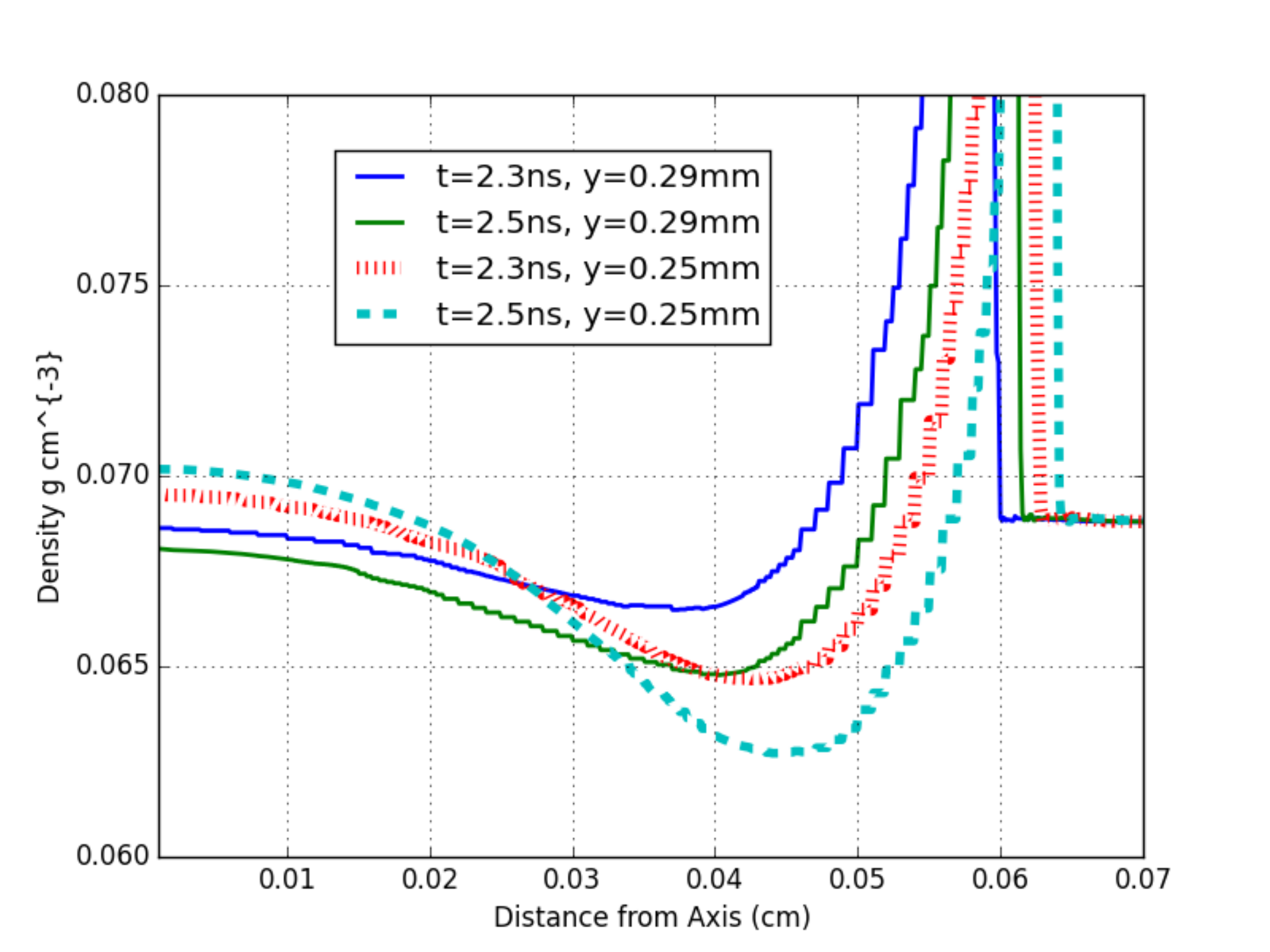}
\caption{Density profile along the line-of-site of the diagnostic
  X-rays at two different positions ($0.025,0.029$\,cm above the
  drive) in our target at 2 different times ($2.3,2.5$\,ns).
  Although, for the most part, the density is flat ($\pm 3$\%) within
  the doped region, there is a density spike near the radiation front.
  The densities measured by a single spectrum can also vary by over
  $\pm 10$\%.}
\label{fig:traceden}
\end{figure}

Figure~\ref{fig:tracetemp} shows the variation in the temperature
distribution across the line of site of the temperature diagnostic as
a function of the position and timing of the diagnostic.  X-rays
produced from a heated krypton source irradiate the target.  The
curvature in the shock front causes large variations across the
target, but we are only concerned with the temperature variations in
the doped foam.  But even in the doped region, the temperature can
vary dramatically (by more than 30\%).  Similarly, the density in the
doped region also has a peak at the shock front
(Fig.~\ref{fig:traceden}).  If the shock front is in the doped region,
the density can vary as much as 10-20\%.  In analyzing this
experiment, we introduce errors if we assume our spectra are produced
by a single density/temperature pair.  However, if we instead use the
full temperature and density structures from the simulations to
produce spectra, we can compare simulated spectra directly to the
diagnostic observation (see Section~\ref{sec:spectra}).  This 
``forward'' analysis approach (using theoretical models to analyze the 
experiment), using the results of 
the simulations to compare to the diagnostic measurements, allows 
us to take advantage of the differences in the temperature profiles 
and test the differences in transport techniques more accurately.

\subsection{Finite Size and Duration of Diagnostic Measurement}

Another factor arguing to use simulated profiles to analyze the data
from the COAX temperature diagnostic is the fact that the spectra are
taken from a slit of finite width (roughly $30 \, \mu m$) and is
measured over a finite time ($\sim 0.1-0.2\,{\rm ns}$).  Across the
slit at the time of the measurement, the temperature varies by more
than $\pm 8$\,eV and the density differences vary by more the $\pm
10$\% (Figures~\ref{fig:tracetemp},\ref{fig:traceden}).  By including
full simulations in our analysis, we can mitigate these errors as well.

\subsection{Modeling Diagnostic Spectra}
\label{sec:spectra}

\begin{figure}[!hbtp]
\centering
\includegraphics[width=2.7in,angle=0]{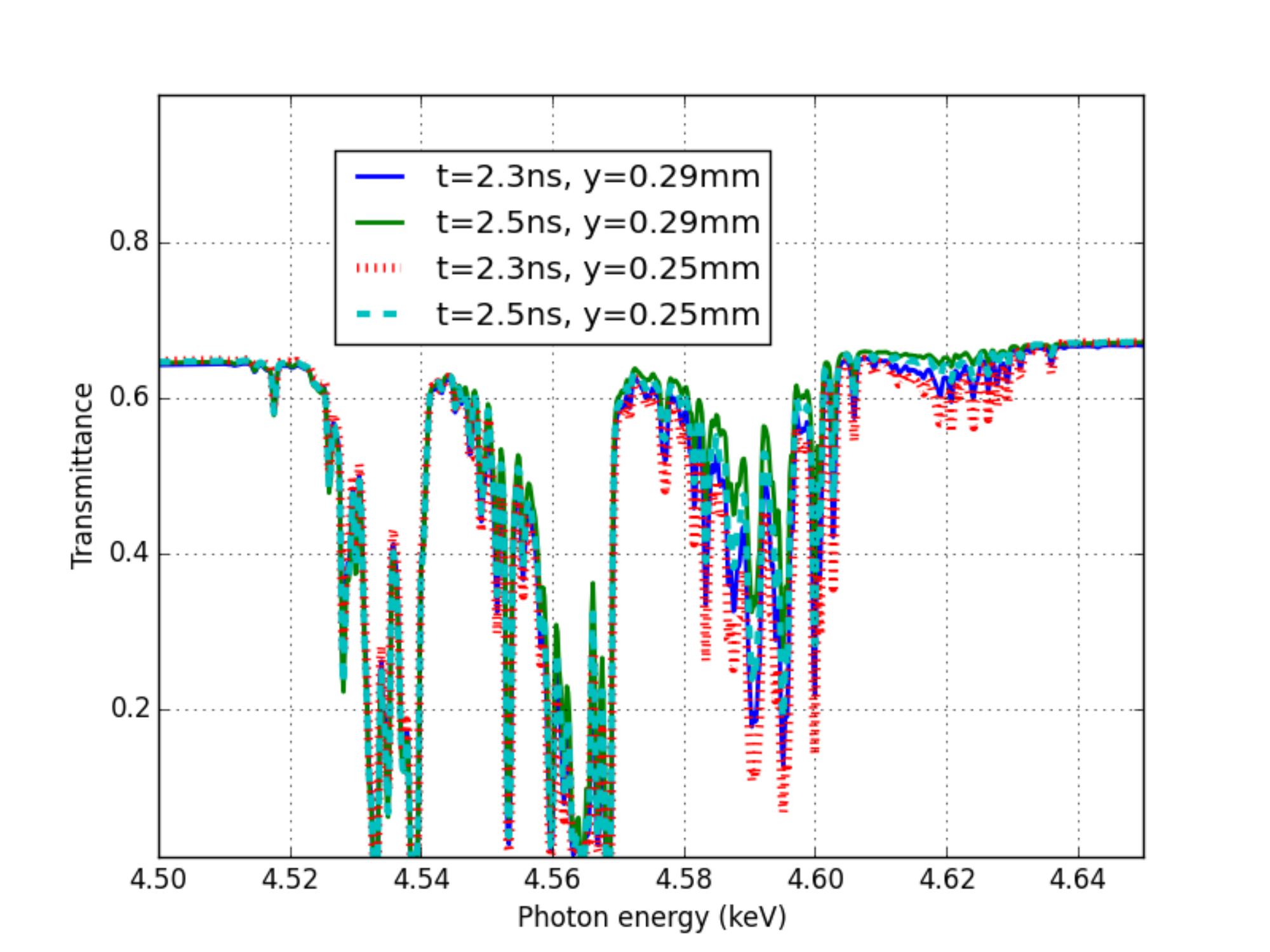}
\caption{Transmission spectra for our temperature diagnostic along a
  series of rays varying the position and timing.  The diagnostic
  actually measures a summation of these spectra because of the width
  of the slit and the duration of the measurement.  Although some of
  the features are strong over all conditions, other features are
  present in some conditions and absent in others.  The sum will
  produce results in between the values shown here.}
\label{fig:spec}
\end{figure}

The variation of temperature and density across the diagnostic
measurement caused both by the curvature in the radiation front as
well as the fact that the diagnostic is not a delta function in width
or time.  As we have discussed, this requires a forward approach where 
we calculate simulated spectra and compare directly to the observations 
(rather than compare inferred temperatures to simulated temperatures).
To calculate these spectra, LANL has
developed a number of ray-trace algorithms (including absorption and
emission) to calculate the transmission from the source.  Ray trace
methods are ideally suited to our calculations.  We can set our
line-out to match the ray between the X-ray backlighter and the detector.
Although our code includes thermal emission from the target, at the
high photon energies of our diagnostic, this emission is negligible.
Hence, the X-ray flux through the target ($L_{\rm detected}(\nu)$) as
a function of frequency ($\nu$) is given simply by
\begin{equation}
L_{\rm detected}(\nu) = L_{\rm source}(\nu) e^{-\tau(\nu)}
\end{equation}
where
\begin{equation}
\tau(\nu)=\int_0^{W} \rho(x) \sigma_\nu(\rho(x),T(x)) dx
\end{equation}
where we integrate along a ray (we assume it is perpendicular to the target so 
the integration is along width of the target, $W$),
$\sigma_\nu(\rho(x),T(x))$ is the total (scattering plus absorption)
opacity in the foam that depends on the composition, density
($\rho(x)$) and temperature ($T(x)$), and $L_{\rm source}(\nu)$ is the
source emission.  Figure~\ref{fig:spec} shows different spectra from
our calculations with the full range of possible spectra within these
combined experimental errors.  Ultimately, combining these
uncertainties with our initial conditions uncertainties argue that we
can only validate the accuracy of our temperature diagnostic to
roughly $\pm 8-10$\,eV.

\section{Next Steps:  Applications of the COAX}
\label{sec:future}

We have reviewed many of the uncertainties in the COAX radiation flow
experiment and its temperature diagnostic.  We confirmed that
equilibration effects in the target do not introduce uncertainties
and, with small enough dopants, there is not a strong stochasticity in
the temperature.  Our simulations also show that the exact angular
distribution and spectrum of the radiation flow do not significantly
alter temperature measurements.

However, we have found that uncertainties in the initial conditions (drive,
foam target) produce temperature variations at the $\pm 8$\,eV level,
even if the position of the shock can be used to constrain these
conditions.  To improve this accuracy, we must measure the drive more
carefully.  Similarly, the width and time duration of temperature
measurement also places a $\pm 8$\,eV error on the temperature
observed.  Combining these errors (assuming in quadrature), this
implies we can validate the temperature diagnostic errors to $\pm
10$\,eV.

Whether or not an experiment like COAX can help constrain our
numerical models depends upon whether we can a) reduce the initial
condition errors and b) design an experiment focused on the
differences in the numerical methods.  Our error analysis thusfar has
modeled the conditions used in the COAX platform development shots.
However, the actual COAX experiment includes plans beyond the basic
platform shots and, by changing the initial conditions, we might
consider the uncertainties in our numerical methods.  We highlight
these future designs here.  The Cassio code is ideally suited to
testing the potential for experiments to test transport methods,
because both the implicit Monte Carlo and $S_N$ methods were
implemented such that the two methods include sources,
etc. identically.  By switching from one method to the other, we can
directly test the differences between transport implementations.

\begin{figure}[!hbtp]
\centering
\includegraphics[width=3.0in,angle=0]{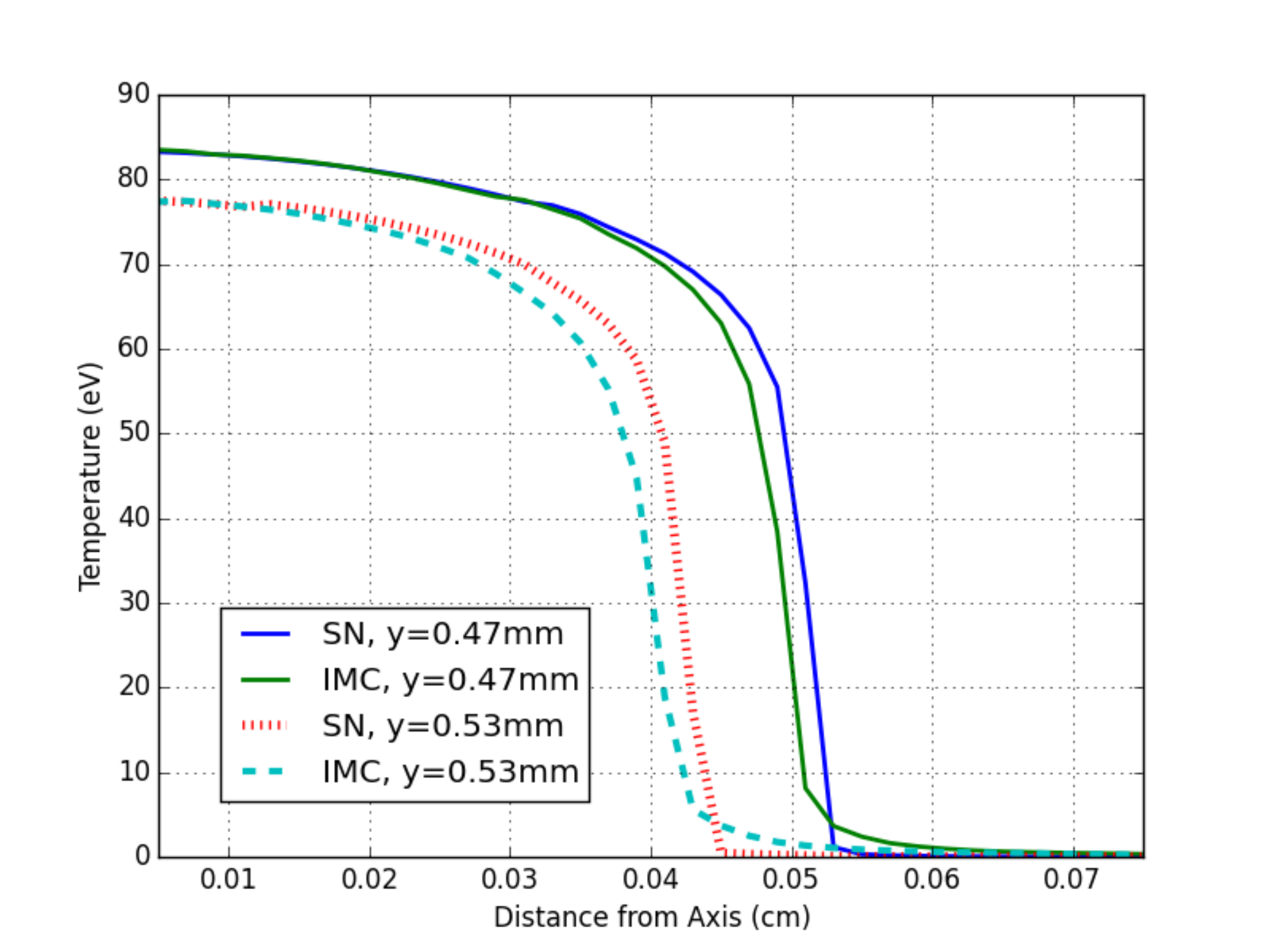}
\caption{Temperature profile along the line-of-site of the diagnostic
  X-rays at two different positions ($0.047,0.053$\,cm above the
  drive) in our target at 2 different times ($3.1,3.3$\,ns) and two
  different drives (solid lines are low drive, dashed lines are high
  drive).  It is possible that the temperature diagnostic, if the
  dopant were deposited in the outer COAX shell would have a
  measureable effect.}
\label{fig:snimc}
\end{figure}

\begin{figure}[!hbtp]
\centering
\includegraphics[width=3.0in,angle=0]{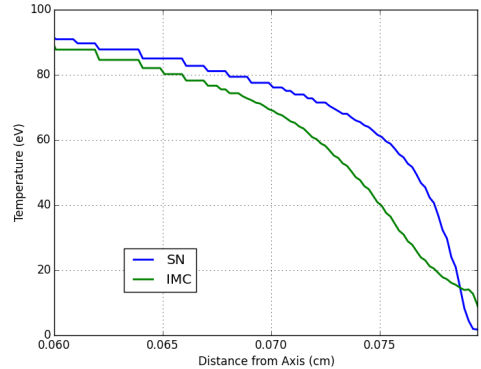}
\includegraphics[width=3.0in,angle=0]{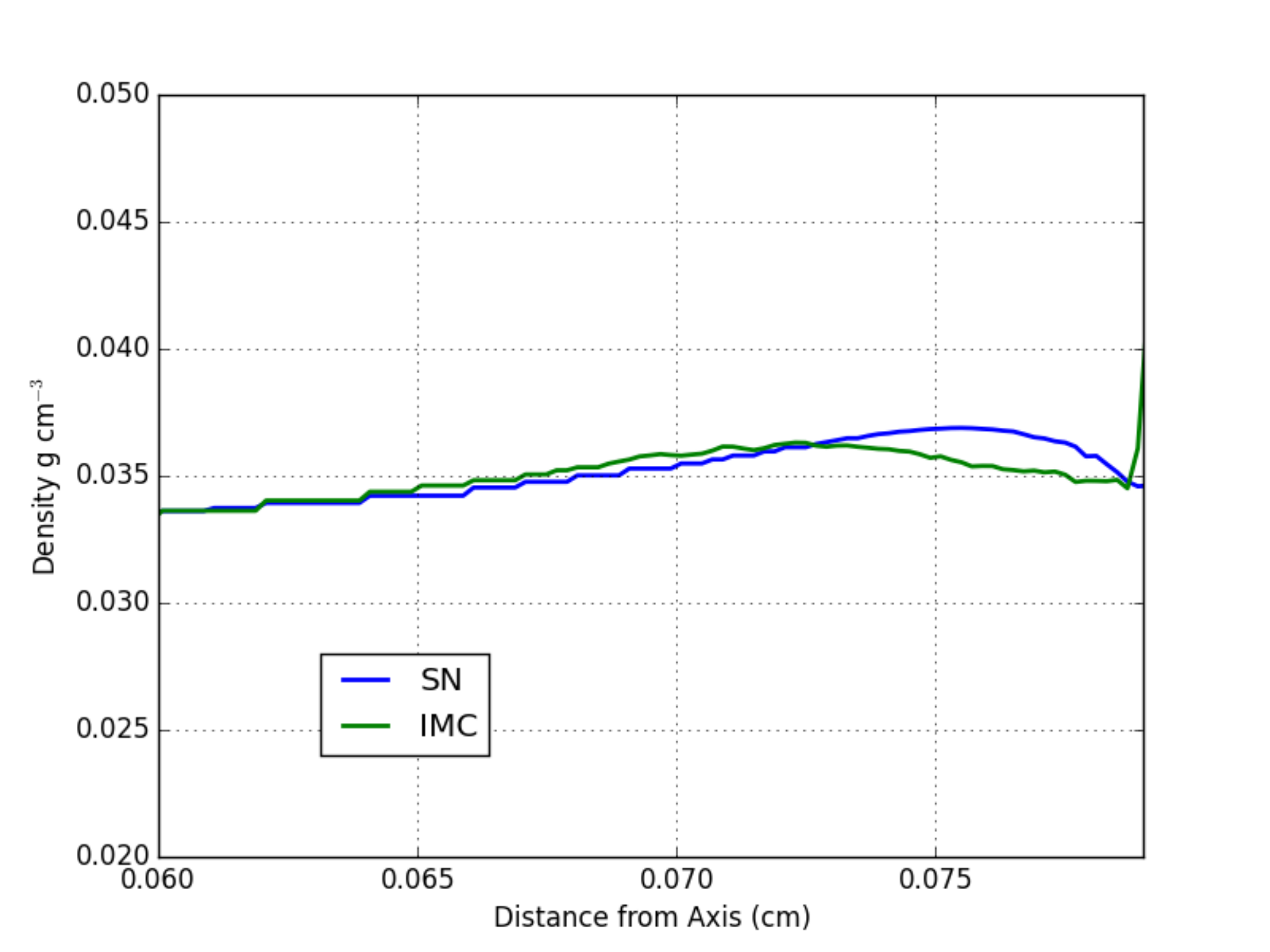}
\caption{Temperature and density profiles along the line-of-site of the diagnostic
for a COAX design where the outer shell density is decreased by a factor of 2.  }
\label{fig:snimcnd}
\end{figure}

\begin{figure}[!hbtp]
\centering
\includegraphics[width=3.0in,angle=0]{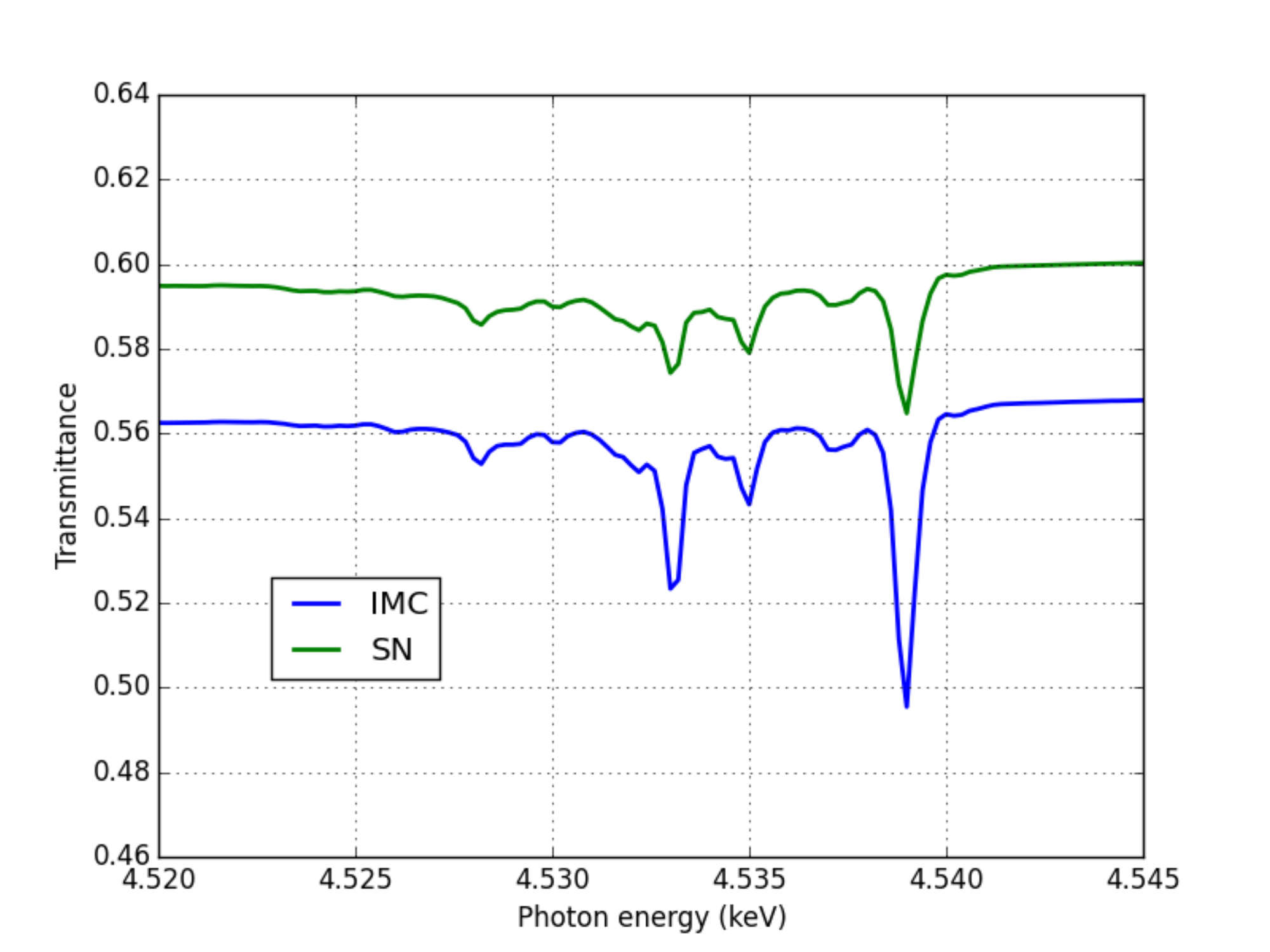}
\caption{Comparison of our spectral diagnostic between the $S_N$ and
  IMC simulations of the doped shell.  The higher temperature in the
  $S_N$ simulation leads to stronger line features and higher over-all
  opacities with respect to the IMC calculation. }

\label{fig:snimcndspec}
\end{figure}

One of the issues with numerical methods in transport is transport
across boundary conditions, and the component COAX platform was
designed to test this boundary layer modeling.  But, to do so, the
diagnostic needs to be in the outer shell.  Figure~\ref{fig:snimc}
compares the temperature along the diagnostic ray-trace for
simulations using implicit Monte Carlo and $S_N$ discrete ordinate
method.  Although the temperature within the doped region from the
platform tests is very similar between these two methods, the amount
of radiation flowing across the boundary layer is more substantial.
The differences are more extreme if we vary the density between the
inner and outer region.  The largest effect occurs if we decrease the
density in the outer layer and this provides a good way to test our
numerical models for this boundary layer physics.
Figure~\ref{fig:snimcnd} shows the density and temperature profile of
the outer shell.  The different transport schemes produce profiles
that have temperatures that vary by up to 20\,eV.  The corresponding
spectral diagnostic with our doped shell are shown in
Figure~\ref{fig:snimcndspec}.  Even with current errors in the
experiment, this difference should be detectable.

\begin{figure}[!hbtp]
\centering
\includegraphics[width=3.0in,angle=0]{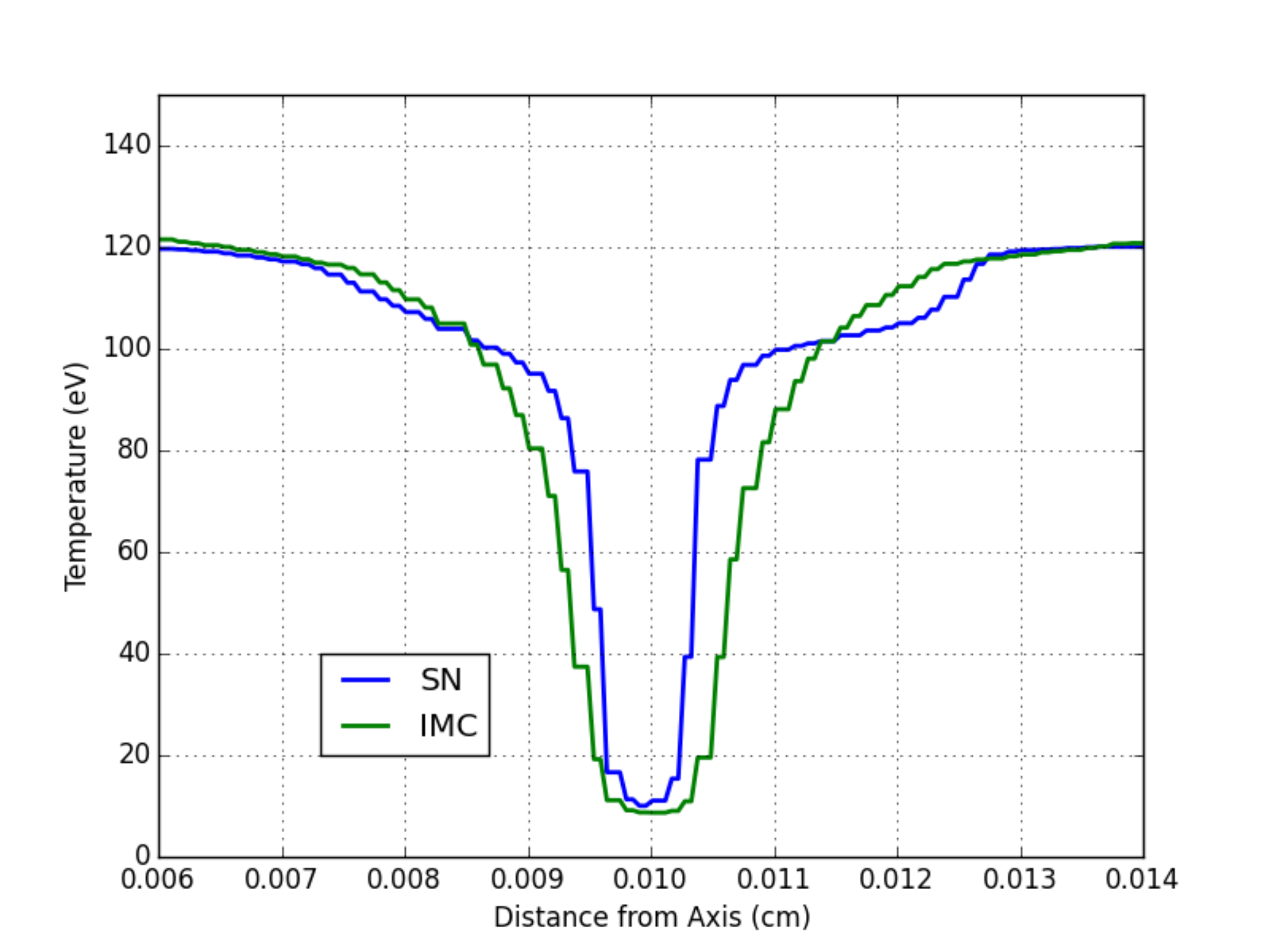}
\includegraphics[width=3.0in,angle=0]{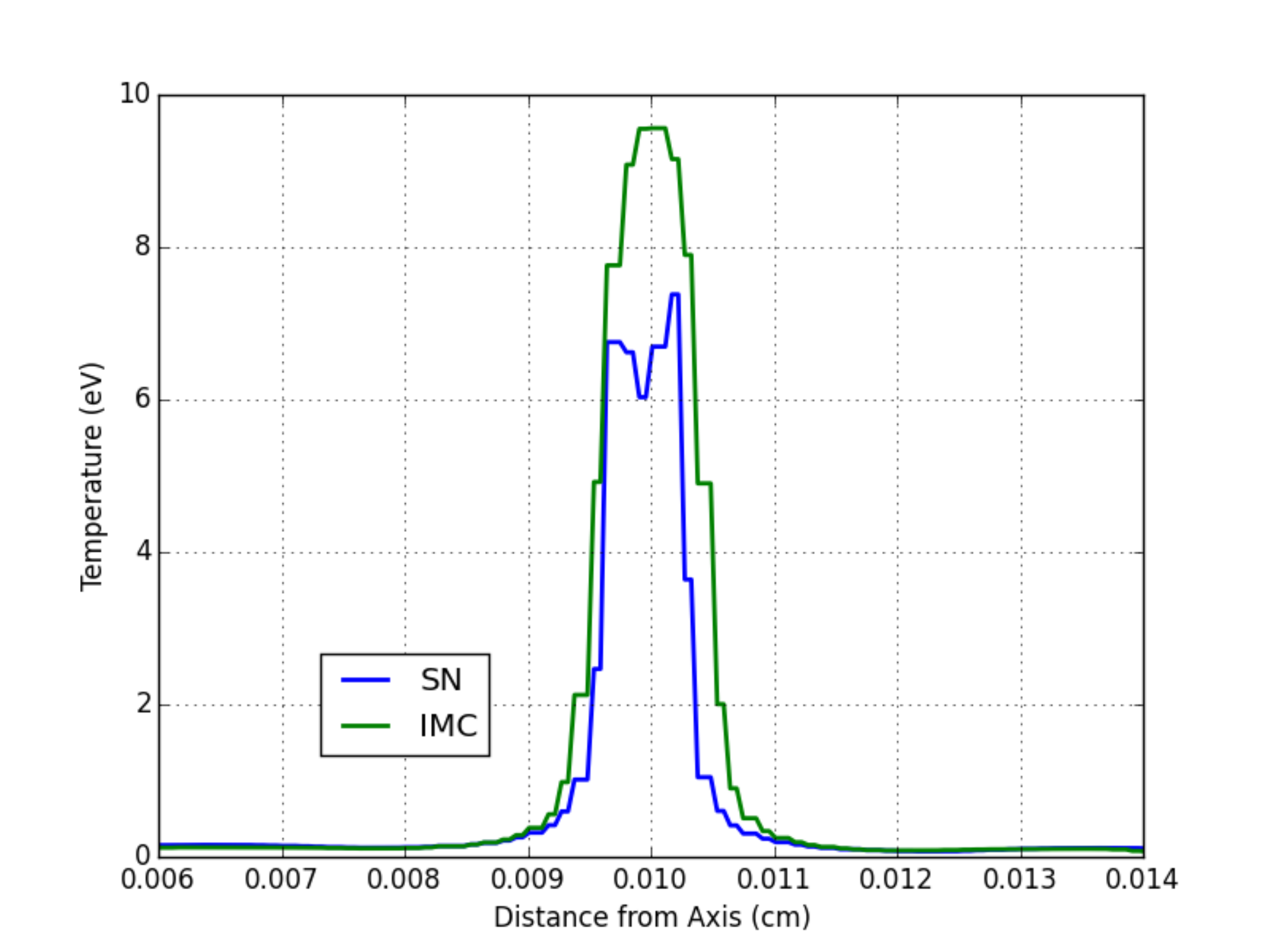}
\caption{Temperature and density profiles along the line-of-site of the diagnostic
across a $10\mu m$ dopant comparing IMC and $S_N$ results.  The foam interface 
is out 0.04\,cm and this image focuses on the diffusion beyond this interface 
where differences in transport methods are most extreme.}
\label{fig:snimcdop}
\end{figure}

One can also broaden how we leverage the COAX platform.  As we
discussed in our error analysis, large dopants take more time to
equilibrate.  This again is an example of radiation flow across a
boundary and our IMC and $S_N$ results are different.
Figure~\ref{fig:snimcdop} shows the density and temperature profiles
across a $10\,\mu m$ dopant 1\,ns after the passage of the shock.  Work 
to design an ideal test case for both Omega and NIF platforms is 
right now underway.

New algorithms are right now being developed to improve the numerics
behind boundary transport in codes.  With better probes of the output
from the hohlraum (underway) and better characterizations of the foam
uncertainties, these future COAX experiments are ideally suited to
testing these new developments.

\section*{Acknowledgements}
This work was performed under U.S. Government contract
89233218CNA000001 for Los Alamos National Laboratory (LANL), which is
operated by Triad National Security, LLC for the U.S. Department of
Energy/National Nuclear Security Administration.

\appendix

\section{Equilibrium Analysis}

To evaluate the electron-ion equilibration times,  we use the Landau-Spitzer (LS) electron-ion rate given by \cite{Spitzer}:
\begin{eqnarray}
\tau_{ei}=\frac{3}{4\sqrt{2\pi}}\frac{m_e m_j c^3}{n_i Z_i^2 e^4 \lambda } \bigg(\frac{k_B T_e}{m_ec^2} +\frac{k_B T_i}{m_i c^2}\bigg)^{3/2}.
\end{eqnarray} 
Here $m_e$ is the electron mass, $m_i$ the ion mass, $k_B$ is the Boltzman constant, $T_e$ is the electron temperature, $T_i$ is the ion temperature, $n_i$ is the ion density, $Z_i$ is the ion charge number, $e$ is the elementary charge, $c$ is the speed of light, and $\lambda$ is the Coulomb logarithm containing details of the collision process. The main issue with this rate is finding accurate Coulomb logarithms. Landau-Spitzer used
\begin{equation}
\lambda =  \ln \bigg(\frac{b_{\rm max}}{b_{\rm min}}\bigg)
\end{equation}
where $b_{{\rm max}}$ and $b_{{\rm min}}$ cutoffs necessary  to prevent the divergences that emerge from their treatment. $b_{{\rm min}}$ is chosen to be a minimum impact parameter consistent with plasma conditions, such as the classical distance of closest approach 
\begin{eqnarray}
b_C = \frac{Z_i e^2 }{ k_B T_i}
\end{eqnarray}
For systems where quantum degeneracy effects are important $b_{{\rm min}}$ is changed to account for quantum diffraction effects by introducing the electron thermal deBroglie wavelength 
\begin{eqnarray}
\Lambda=\sqrt{\frac{\hbar^2}{4 m_e k_B T_e}}
\end{eqnarray}
where $\hbar$ is the Planck constant, and $b_{\rm max}$  is chosen to be a screening length occurring from collective plasma effects. A large number of formulae for the Coulomb logarithm exist in the literature. We choose here the expression proposed by Gericke, Murillo, and Schlanges  (GMS) \cite{GMS}. Their formula was successfully validated against molecular dynamics simulations results across a wide range of temperature and density \cite{Graziani,GlosliPRE08}. GMS suggested an effective Coulomb logarithm as
\begin{eqnarray}
\lambda=   0.5\ln\bigg[1+\frac{\lambda_D^2+a_i^2}{\Lambda^2+b_C^2}\bigg]
\label{eq:CL}
\end{eqnarray}
where the ion sphere radius is given in terms of the ion density $n$ as
\begin{eqnarray}
a_i = \bigg(\frac{3}{4\pi n}\bigg)^{1/3}
\end{eqnarray}

\subsection{Timescales in the COAX Experiment}

For a pure carbon foam with density  $\rho = 60 $ mg cm$^{-3}$ and temperature $150$ eV, and electron temperature of $T_e=100$
With the above formula, we estimated the electron-ion relaxation times for  a TiSi$_5$O$_{12}$ foam target at temperature of $100$ eV for near-vacuum densities ($\rho =0.03 $ mg cm$^{-3}$) and higher density $\rho = 60 $ mg cm$^{-3}$ and for an electron temperature of $100$ eV. The LS calculations show that the average foam-ion takes around $\sim 300$ ns to equilibrate with the electrons while for the high density case  the equilibration time is around $\sim 0.3$ ns.
Figures 1 and 2 display the electron-foam relaxation predictions from Landau-Spitzer as function of time for two different  holhraum densities and temperature.

\begin{figure}[!b]
  \centering
  \begin{minipage}[b]{0.45\textwidth}
    \includegraphics[width=1\textwidth]{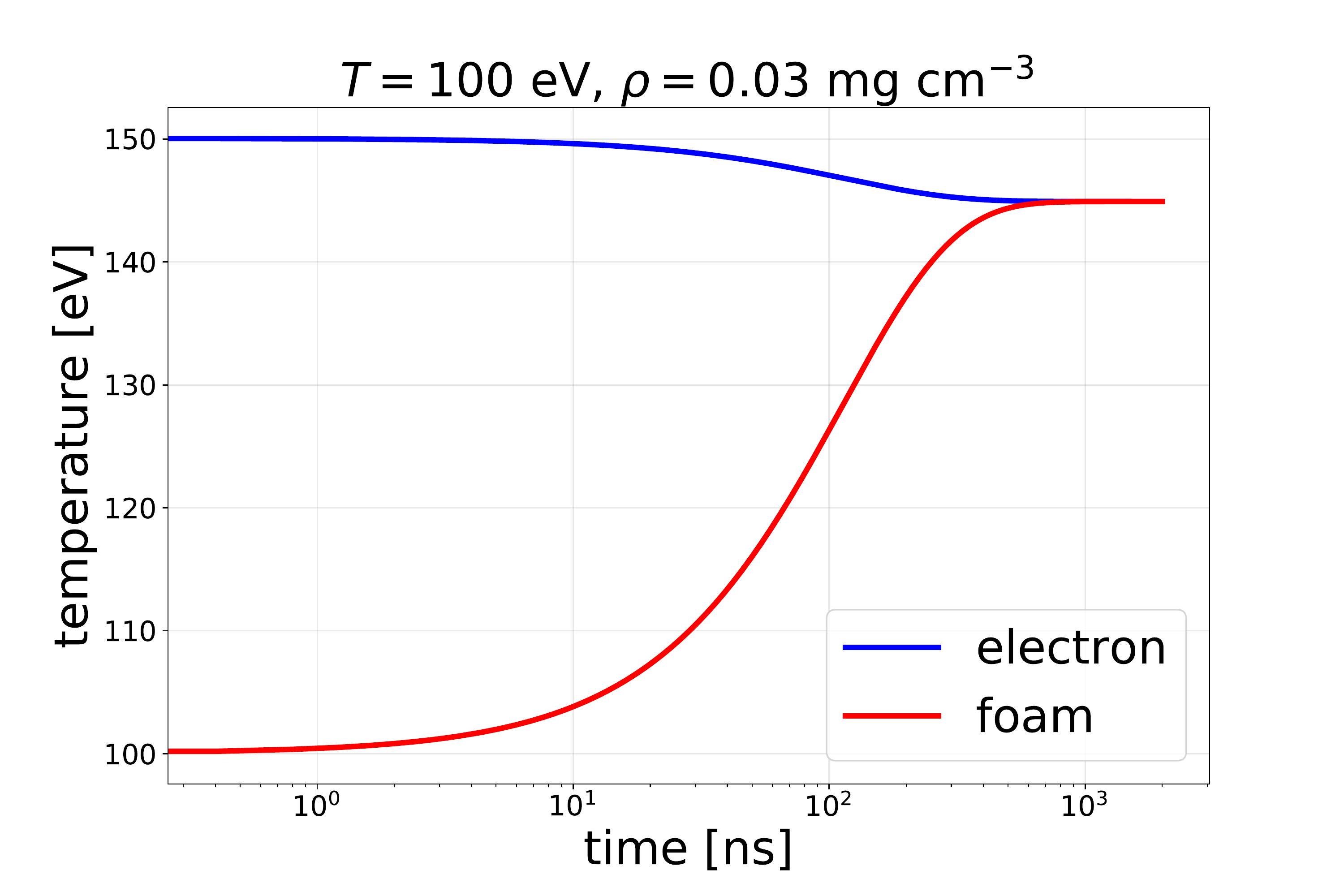} 
    \caption{Electron  top curves  and foam  bottom curves  temperature relaxation is shown based on Landa-Spitzer for a density $\rho=0.03$ mg cm$^{-3}$.}
    \label{fig:1}
  \end{minipage}
  \hfill
  \begin{minipage}[!b]{0.45\textwidth}
    \includegraphics[width=1\textwidth]{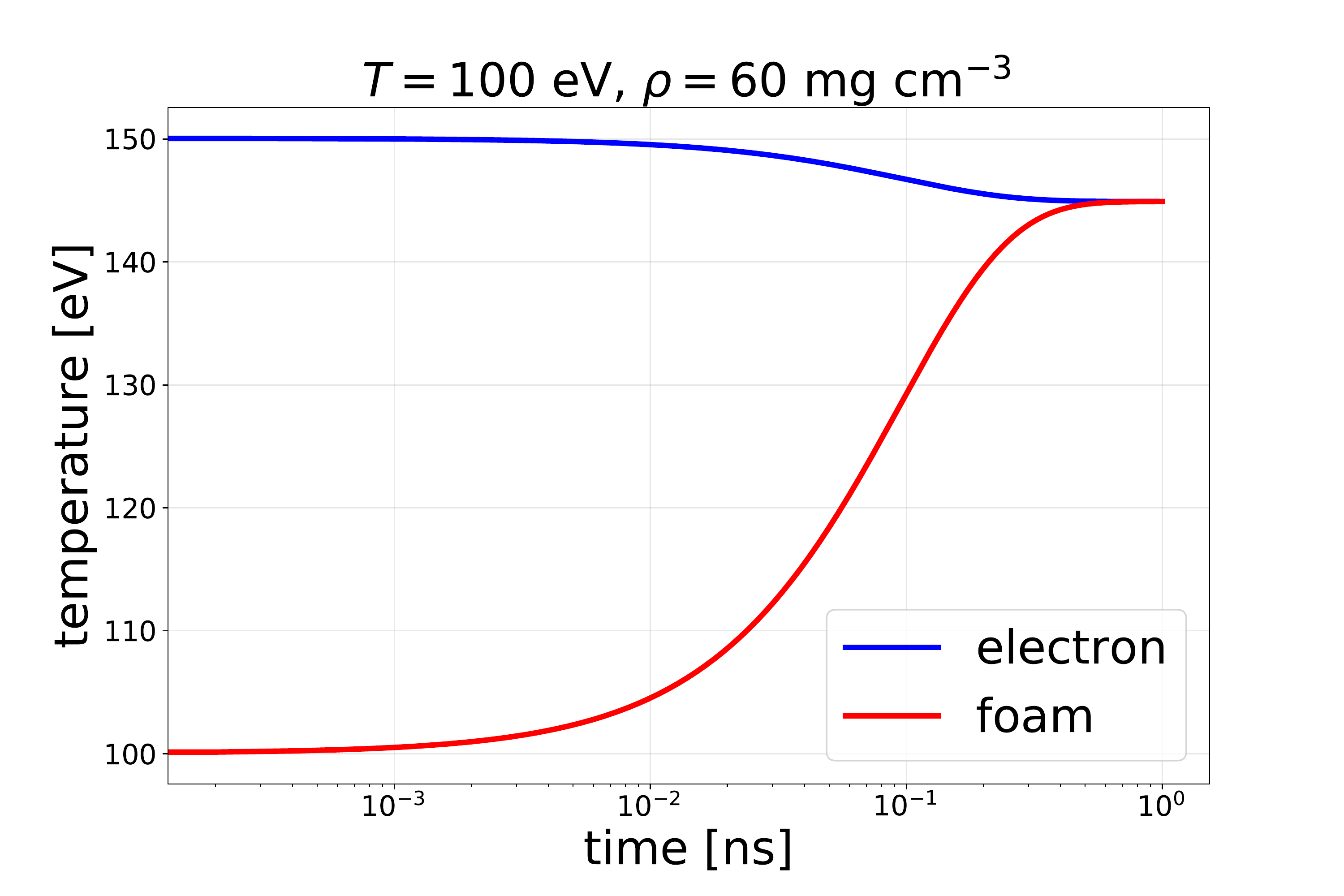}
    \caption{Same as Fig. \ref{fig:1} with density $\rho= 60$ mg cm$^{-3}$.}
  \end{minipage}
\end{figure}

\subsection{Maxwellinization of Electrons distribution function}

Plasmas produced in high energy density experiments are often characterized by steep temperature and density gradients.
Because of these gradients, the thermal mean-free path of the electron ($\lambda_{\rm mfp}$) becomes large with respect to the characteristic scale length of the temperature $l_T= |T/\nabla T|$. Electrons with mean-free path collisional larger than the scale length of the  temperature can escape gradients before being scattered and depositing their energy into the plasma, leading to a distortion of the electrons  distribution function (EEDF) away from a Maxwellian distribution \cite{Bell83, LMV, Albritton86}.  
This non-Maxwellian behavior of the  EEDF shows the nonequilibrium behavior of the electron plasma, therefore suggesting the need for a kinetic description for the electrons.

One consequence of the distortions of the EEDF, within hydrodynamics codes, is that  the classical Spitzer-Harm formula for the heat-flux can exceed the heat flux of free streaming electrons \cite{2011HEDP....7..180R}. Historically, through analysis of experimental results, interial confinement fusion (ICF) designers very quickly realized this problem of over-estatimation of the heat flux, and decided to use flux limiter models (which cannot be predictive, and need to be calibrated against experiments), and  later nonlocal convolution kernel to represent the nonequilibrium effects \cite{LMV,SNB,MCG}.

Recent studies, however, have suggested that for problems relevant to ICF hohlraum conditions, the nonlocal models depart strongly from Vlasov-Fokker-Planck simulations \cite{Brodrick17,Sherlock17}. These observations indicate the need to go beyond equilibrium and near-equilibrium models and adopt kinetic approaches (Boltzmann, Vlasov-Fokker-Planck)  to describing electrons dynamics for the near-vacuum hohlraum experiments. Similarly, in these near-vacuum hohlraum experiments ions can also displays non-Maxwellian effects as have been discussed recently in Refs. \cite{2018PPCF...60f4001R,2017NucFu..57f6014R}

\bibliographystyle{plainnat}
\bibliography{master}

\end{document}